Yong-Lei Wang, Sten Sarman, Mikhail Golets, Francesca Mocci, Zhong-Yuan Lu and Aatto Laaksonen

# 4 Multigranular modeling of ionic liquids

**Abstract:** Ionic liquids (ILs) are a special category of molten salts with melting points near ambient temperatures (or by convention below 100 °C). Owing to their numerous valuable physicochemical properties as bulk liquids, solvents, at surfaces and in confined environments, ILs have attracted increasing attention in both academic and industrial communities in a variety of application areas involving physics, chemistry, material science and engineering. Due to their nearly limitless number of combinations of cation–anion pairs and mixtures with cosolvents, a molecular level understanding of their hierarchical structures and dynamics, requiring strategies to connect several length and time scales, is of crucial significance for rational design of ILs with desired properties, and thereafter refining their functional performance in applications. As an invaluable compliment to experiments from synthesis to characterization, computational modeling and simulations have significantly increased our understanding on how physicochemical and structural properties of ILs can be controlled by their underlying chemical and molecular structures. In this chapter, we will give examples from our own modeling work based on selected IL systems, with focus on imidazolium-based and tetraalkylphosphonium-orthoborate ILs, studied at several spatio-temporal scales in different environments and with particular attention to applications of high technological interest. We start by describing studies performed using *ab initio* methods on force field development for tetraalkylphosphonium-orthoborate ILs, and computational studies on thermal decomposition of these ILs. The delicate interplay between hydrogen bonding and π-type interactions in an imidazolium-orthoborate IL was studied by performing *ab initio* molecular dynamics simulations. On the atomistic level, atomistic simulations were performed with constructed force field parameters to study intrinsic molecular interactions between residual water molecules and tetraalkylphosphonium-orthoborate ionic species. For a typical trihexyltetradecylphosphonium bis(oxalato) borate IL at varied concentrations, microstructures and dynamics were systematically analyzed as water concentration increases. The liquid viscosities of typical trihexyltetradecylphosphonium-based ILs were estimated through equilibrium atomistic simulations using Green–Kubo relation with charge scaling factors on ionic species.

**Yong-Lei Wang,** Department of Materials and Environmental Chemistry, Stockholm University, Sweden; Department of Chemistry, Stanford University, The United States
**Sten Sarman, Aatto Laaksonen,** Department of Materials and Environmental Chemistry, Stockholm University, Sweden
**Mikhail Golets,** AkzoNobel Surface Chemistry AB, Stenungsund, Sweden
**Francesca Mocci,** Department of Chemical and Geological Sciences, University of Cagliari, Italy
**Zhong-Yuan Lu,** Institute of Theoretical Chemistry, Jilin University, China

Additionally, atomistic simulations were combined with X-ray scattering experiments to investigate phase behavior and ionic structures in IL mixtures with varied concentrations. Peculiar features of X-ray scattering spectra of IL mixtures with organic solvent are also discussed with an example dealing with ethyl ammonium nitrate and acetonitrile mixtures. On the mesoscopic level, a united-atom model for trihexyltetradecylphosphonium cation and coarse-grained models for butylmethy-limidazolium hexafluorophosphate were proposed, and effective interactions between coarse-grained beads were validated against experimental and computational data. Concluding remarks on multiscale strategies in understanding and predictive capabilities of ILs and IL mixtures are addressed in the final section. An outlook is provided to highlight future challenges and opportunities associated with IL materials in multiscale modeling community.

## 4.1 Introduction

Ionic liquids (ILs) refer to a special category of molten salts, entirely consisting of complex, sterically mismatched molecular ions with their melting points at or close to room temperature [1–3]. The first room temperature IL was reported already more than 100 years ago by Pauls Valdens (Paul Walden) who presented the synthesis of ethylammonium nitrate (EAN) by neutralizing ethylamine with concentrated nitric acid in 1914 [4]. EAN is clear, colorless, odorless and has a melting point of 13–14 °C and a rather low viscosity [4, 5]. However, this early report did not receive much attention from the scientific community, and it was not foreseen that such salt materials would become of widespread interest decades later. In 1951, Hurley and Wier reported the synthesis of organic chloroaluminates by mixing alkylpyridinium chlorides with aluminium compounds, which is now considered as the first generation of ILs [6]. Unfortunately, these organic chloroaluminates are not stable in the presence of moisture because of their rapid hydrolysis. Additionally, their acidity and basicity are not easy to regulate [6, 7], and more detailed studies on these compounds started from 1970s [2, 8]. In 1992, Wilkes and Zaworotko prepared moisture/water-stable ILs consisting of imidazolium cations and tetrafluoroborate ([BF$_4$]) anion [9, 10]. It became clear that many ion combinations could form air- and water-stable ILs [2, 10–12]. Immediate scientific research on synthesis, characterizations and applications of ILs got an upswing in academia and industry communities [12–21].

The most frequently studied cationic structures have organic moieties, such as imidazolium, guanidinium, morpholinium, piperidinium, pyrazolium, pyridinium, pyrrolidinium, thiazolium, sulfonium, tetraalkylammonium and tetraalkyl-phosphonium ions. The anionic parts can be either organic or inorganic entities including acetate, halogens, hexafluorophosphate ([PF$_6$]), [BF$_4$], orthoborate, nitrate ([NO$_3$]), alkylsulfonate, alkylsulfate, bis(trifluoromethanesulfonyl)imide ([NTF$_2$]), alkylphosphate, trifluoromethylsulfonate ([TFO]), etc. [1, 8, 12–14, 16, 22, 23]. ILs have remarkable and multifaceted physicochemical characteristics, such as negligible volatility, low flammability, reasonable viscosity-conductivity feature, acceptable biocompatibility, high thermal-oxidative stability, wide electrochemical window, as well as outstanding ability to dissolve solutes of diverse polarities [1, 3, 7, 22, 24, 25]. An additional feature of ILs is that their physicochemical properties and microstructural organization can be widely tuned in a controllable fashion through judicious combinations of different cation–anion moieties in a general way, and by mutating specific atoms in constituent cations or anions [3, 22, 25, 26]. These characteristics make ILs exceptionally attractive and reliable candidates as environmentally benign alternative to conventional molecular solvents in synthetic chemistry to offer precise control over growth rate, particle size and morphology of nanomaterials [8, 11, 18, 23]; useful reaction media in catalytic chemistry to optimize yield, selectivity, substrate solubility, product separation and enantioselectivity [8, 11, 18, 27]; valuable working fluids in separation technology through selective absorption of gas molecules [13, 16, 19, 23, 28]; promising versatile electrolytes in electrochemical energy devices with tunable electrical conductivities and varied electrochemical stability windows [7, 13, 17, 20, 21, 29–31]; suitable lubricants in tribology to reduce wear and frictions between solid sliding contacts [14, 15, 32, 33].

Due to the existence of an enormous number of possible cation–anion combinations, there are also nearly limitless opportunities to produce neat ILs and their multicomponent mixtures with distinct molecular structures and physical properties [3, 22, 23]. However, it is not feasible to systematically pick up ion combinations to synthetize, purify and characterize them as they are simply too many. An efficient and reliable predictive tool is needed to obtain a molecular level understanding of the many competing intermolecular interactions between and within ionic species to guide the development cycle by providing expedient predictions of physicochemical properties for neat ILs and IL mixtures as well as providing an improved fundamental understanding of ILs and thereby obtain data to develop fast structure–property relationship models [2, 25].

Computer simulations, in close interplay with experiments, can provide much fundamental understanding of complicated phenomena on molecular level, which is particularly useful for ILs because of their large diversities and their complicated landscape of interactions. Various simulation methods are available to use at different spatio-temporal scales depending on specific targets requested from studied model IL systems [2, 31, 34–48]. Multiscale modeling approach generally involves several simulation techniques, including quantum chemical calculations, *ab initio*, atomistic and coarse-grained (CG) molecular dynamics simulations, to study specific model systems. The advantages and drawbacks of various methods in different spatio-temporal scales coexist and are often interchangeable [35, 42–45]. For quantum chemical methods, that is, wave-function-based (Hartree–Fock) calculations or density functional theory (DFT), the detailed physical insight into electronic structures of ILs is limited to small systems containing only a few ion pairs. The advantage of these quantum chemical methods stems from their inherent accuracies through the electronic structure and can be systematically improved until predetermined target accuracy is achieved but on an increasing cost of computing time. Static properties of molecular clusters can be typically calculated even if some of these properties are not exclusively representatives of bulk systems [38, 39, 49–51]. Using DFT-based (Born–Oppenheimer or Car–Parrinello) molecular dynamics simulations it is now possible to set up a system up to 100 ion pairs and cover a time scale of a few nanoseconds. However, this is still far from being sufficient to sample any dynamical properties of ILs as they require much bigger systems and simulation times of hundreds of nanoseconds.

To reach much longer time scales, classical atomistic simulations are carried out to obtain dynamical and transport properties of ILs while treating systems consisting of hundreds of thousands of atoms and simulating over nano or even microsecond time scales [24, 34, 35, 41–45, 47, 52, 53]. Since the accuracy of thermodynamics, microstructures and transport properties of model systems depend ultimately on appropriate force field parameters, one can iteratively fine-tune interaction parameters and validate them at each refinement stage through underlying *first-principles* calculations and comparison of computed properties against experimental data when available. Additionally, one can further construct CG models allowing for large systems to be simulated for long times, thus revealing liquid structural and dynamical features of IL systems that are difficult to predict through conventional atomistic simulations [31, 35, 36, 41–46, 54–56].

In this chapter, we will highlight our work done during the lifetime of the COST action EXIL. We have studied many systems on butylmethylimidazolium- ([BMIM]), EAN- and tetraalkylphosphonium-orthoborate-based ILs. Typical molecular structures of these studied ILs are shown in Figure 4.1. We did start with force field development and validation for tetraalkylphosphonium-orthoborate ILs. Quantum chemical calculations were performed to study thermal decomposition mechanism of typical tetraalkylphosphonium-orthoborate ILs. *Ab initio* molecular dynamics (AIMD) simulations were performed to investigate the complex interplay between hydrogen bonding (HB) characteristics and π-type stacking features in dimethylimidazolium bis(oxalato) borate ([MMIM][BOB]) IL. With constructed force field parameters, atomistic simulations were performed to study thermodynamics, microstructural and dynamical properties of tetraalkylphosphonium-orthoborate ILs, as well as the dependence of these properties on residual water contents in typical trihexyltetradecylphosphonium ([$P_{6,6,6,14}$])-orthoborate ILs. We wanted to find out why it is difficult to remove residual water from a hygroscopic IL by studying four [$P_{6,6,6,14}$]-orthoborate ILs through calculations of solvation free energy of dissolving water from gas phase into bulk ILs and followed the coordination of dissolved water molecules among ions in IL matrices. Atomistic simulations were carried out to elucidate microscopic interfacial structures and ordering arrangement of [BMIM]-based ILs and [$P_{6,6,6,14}$] bis(mandelato)borate ([BMB]) IL in confined environment characterized by different surface charge densities. The liquid viscosities and rheological properties of [$P_{6,6,6,14}$]-orthoborate ILs were computed using nonequilibrium shear flow simulations and equilibrium atomistic simulations using Green–Kubo relations. Atomistic simulations were combined with X-ray scattering technique to reveal striking scattering patterns of various IL mixtures. Additionally, we developed CG models for [$P_{6,6,6,14}$] cation and [BMIM][$PF_6$] IL to perform simulations at extended spatio-temporal scales. In addition to a benefit of computational efficiency, it is expected that these proposed CG models can reveal essential structural and dynamical properties of ILs at mesoscopic level by integrating over less important degrees of freedom at atomic level.

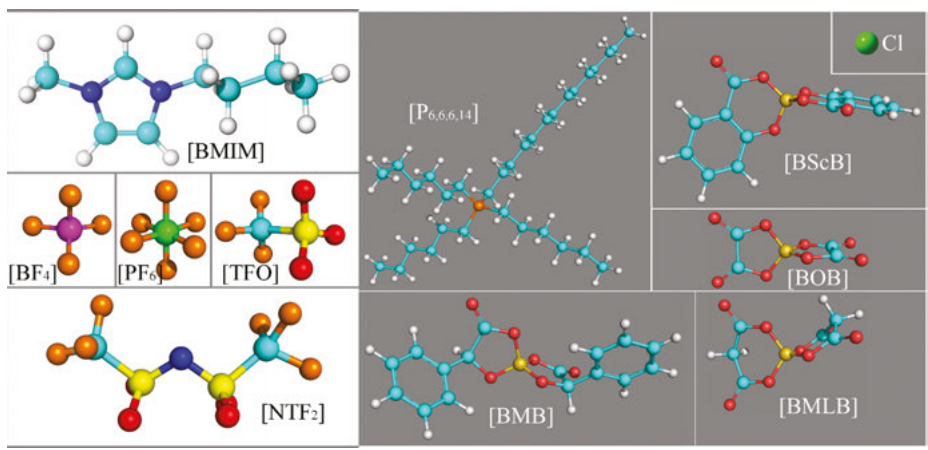

**Figure 4.1:** Representative ion structures of cations and anions discussed in this contribution.

## 4.2 Quantum chemical calculations

### 4.2.1 Force field development

*First-principles* quantum chemical calculations and *ab initio* molecular dynamics simulations of ILs can provide accurate microstructural information without any experimental or empirical input. However, being computationally demanding, these methods are mainly applied to study chemical reactions and to derive effective interactions between ion pairs and to determine their delicate interactions with solute molecules in small systems and at short time scales [38, 39, 46, 57].

Since 2000, there have been continuous efforts to develop and refine force field parameters for IL systems. The first force field development for [MMIM] and ethylmethylimidazolium ([EMIM]) cations coupled with chloride (Cl) and [$PF_6$] anions were carried out by Hanke and coworkers [58]. A Buckingham repulsion dispersion term was included in nonbonded repulsion–dispersion interactions between atomic sites in cationic models to reproduce experimental crystal structures. Subsequent atomistic simulations were performed at high temperatures due to the high melting points of [MMIM]Cl and [MMIM][PF$_6$] ILs and limited computational resources. In a following study, the excess chemical potentials of hydroxyl, ether and alkane solutes in these ILs were calclated using thermodynamic integration method [59]. This is the first example of free energy calculation within an IL system, and it confirmed the importance of HB and charge–charge interactions for the solvation behavior of ILs.

These early works were followed by several groups who came up with force field developments either for specific ILs [24, 34, 60–64] or for an entire class of ILs [52, 53, 65–69]. These proposed interaction parameters for ILs were established by extending and refining the well-developed force fields, such as AMBER, CHARMM, OPLS and GROMOS, and thereafter were validated against available experimental properties, including liquid densities, spectroscopic, neutron scattering and diffraction data and X-ray crystallographic data on imidazolium-, pyridinium-, tetraalkylammonium- and guanidinium-based ILs.

Following a similar procedure, we developed an atomistic force field [70] for a new class of tetraalkylphosphonium-orthoborated ILs observed to have high friction-reducing and antiwear properties in tribological applications [33]. This force field is based on the AMBER framework employing the following functional forms to describe intra- and inter-molecular interactions between ionic species:

$$U_{total} = \sum_{bonds} K_r (r - r_{eq})^2 + \sum_{angles} K_\theta (\theta - \theta_{eq})^2 + \sum_{dihedrals} K_\phi 2[1 + \cos(n\phi - \gamma)]$$
$$+ \sum_{i<j} \left\{ 4\varepsilon_{ij} \left[ (\sigma_{ij} r_{ij})^{12} - (\sigma_{ij} r_{ij})^6 \right] + q_i q_j 4\pi\varepsilon_0 \varepsilon_r r_{ij} \right\}$$

The first three terms represent the bonded interactions, that is, (harmonic) bond and angle, and dihedral potentials, and corresponding potential parameters have their usual meaning. The nonbonded interactions are described in the last term, including van der Waals (vdW, here in the Lennard-Jones 12-6 form) and Coulombic interactions between atom-centered point charges. The vdW interaction parameters between different atoms are obtained from the Lorentz-Berthelot combining rules with $\varepsilon_{ij} = \sqrt{\varepsilon_{ii}\varepsilon_{jj}}$ and $\sigma_{ij} = (\sigma_{ii} + \sigma_{jj})/2$.

In the force field development, the optimized molecular geometries of isolated tetraalkylphosphonium cations, orthoborate anions, as well as the bounded tetraalkylphosphonium-orthoborated ion pair structures were obtained from quantum chemical calculations. The optimized tributyloctylphosphonium ([P$_{4,4,4,8}$]) [BOB] ion pair is characterized by a piggy-back structure with [BOB] anion riding on octyl chain in [P$_{4,4,4,8}$] cation, as shown in Figure 4.2. Atomic partial charges were obtained using standard restraint electrostatic potential methodology to fit molecular electrostatic potential generated from *ab initio* calculations. The vdW parameters were taken from the AMBER force field, and the bond stretching and angle bending force constants available in AMBER framework were subsequently adjusted to reproduce vibration frequency data derived from both experimental measurements and *ab initio* calculations. The missing bond, angle and dihedral terms in AMBER framework were obtained by fitting torsion energy profiles deduced from *ab initio* calculations.

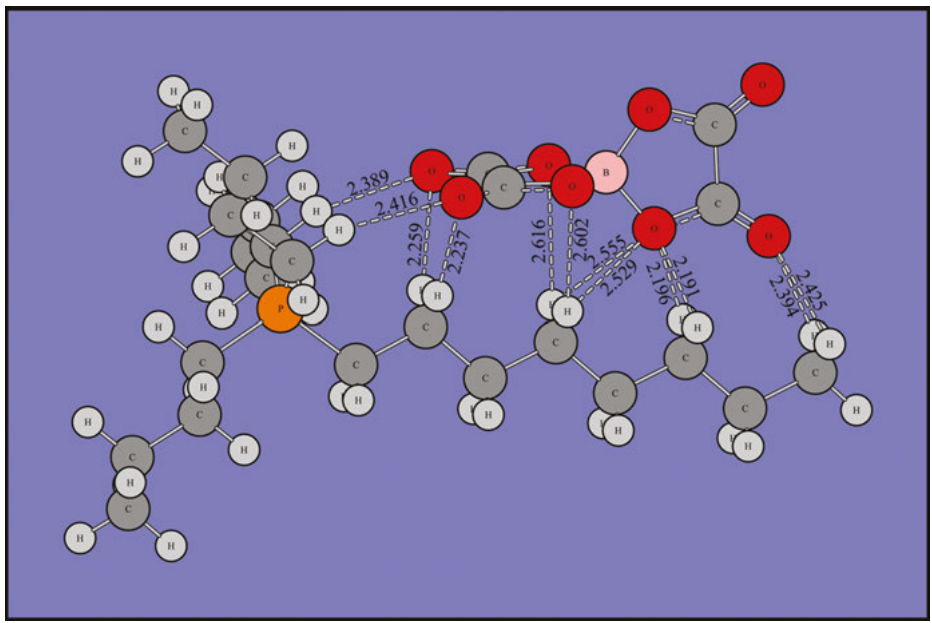

**Figure 4.2:** Optimized [P$_{4,4,4,8}$][BOB] ion pair structure obtained from quantum chemical calculations at B3LYP/6-311++G(d) level of theory. The unit of O...H distance is Å.

In order to validate the proposed force field parameters, extensive atomistic simulations were performed for 12 tetraalkylphosphonium-orthoborate ILs [70]. The predicted densities for neat ILs and the [P$_{6,6,6,14}$][BOB] sample with a water content of approximately 2.3–2.5 wt% are in excellent agreement with available experimental data [33]. The calculated diffusion coefficients of tetraalkylphosphonium cations and orthoborate anions are qualitatively consistent with available experimental viscosity data. The spatial distributions of boron (B) and oxygen (O) in four orthoborate anions ([BOB], bis(malonato)borate ([BMLB]), [BMB] and bis(salicylato)borate ([BScB])) around a [P$_{4,4,4,8}$] cation, as shown in Figure 4.3, indicate that there are mainly four high probability domains for orthoborate anions in coordinating a [P$_{4,4,4,8}$] cation in the first solvation shell. The B atom in [BOB] anion exhibits dispersed tetrahedral distributions around a central [P$_{4,4,4,8}$] cation. An increase in anionic group size from [BOB] to [BMLB], [BMB] and [BScB] leads to the expansion of spatial distribution domains for B atoms (meshed yellow contour surfaces) around [P$_{4,4,4,8}$] cation. The spatial distributions of O atoms (solid red contour surfaces) in orthoborate anions are characterized by trefoil-like structures, and follow a similar tendency as those for boron atoms as anionic size increases.

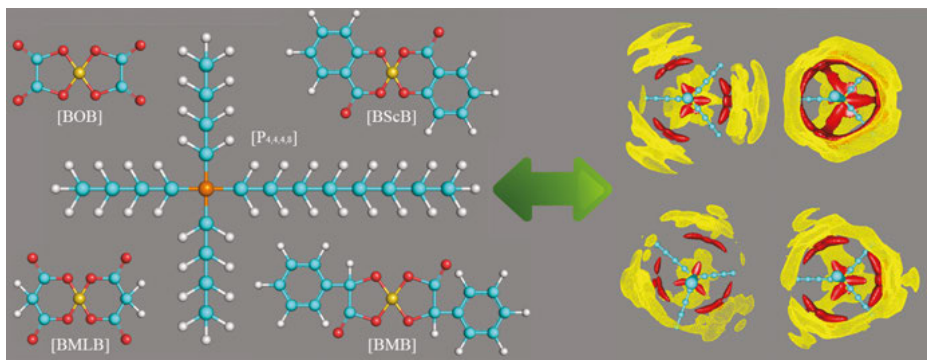

**Figure 4.3:** Spatial density distribution functions of boron atom (meshed yellow surface) and oxygen atoms (solid red surface) in four orthoborate anions around a $[P_{4,4,4,8}]$ cation obtained from atomistic simulations. The yellow and red contour surfaces correspond to 4.0 and 5.5 times of the average number densities of the corresponding atoms in bulk system.

## 4.2.3 Quantum chemical calculations on ionic liquid thermal decomposition

Understanding thermal decomposition of ILs on the nascent metal surface is important for the explanation of lubrication mechanisms in the presence of IL lubricants. In a recent work, a combined quantum chemical modeling and experimental approach was utilized to explain thermal decomposition of $[P_{4,4,4,8}][BOB]$ and $[P_{4,4,4,8}]Cl$ ILs [57]. Quantum chemical calculations (i.e., vibration analysis and potential energy scans) revealed ionic structural changes of $[P_{4,4,4,8}]$ cation and [BOB] anion during thermal decomposition. The $[P_{4,4,4,8}]$ cation is rigid due to its stable central polar segments, and [BOB] anion exhibits high structural and energetic symmetry properties. The cleavage of B—O bond in [BOB] anion initiates the thermal decomposition in neat $[P_{4,4,4,8}][BOB]$ IL, as shown in Figure 4.4. The activation barrier in the beginning of reaction is 246.3 kJ mol$^{-1}$. The resulting monocyclic anions further react with $[P_{4,4,4,8}]$ cation in which P—O bonds are least stable. Three butyl chains in $[P_{4,4,4,8}]$ cation are oriented toward [BOB] anion with activation barriers ranging from 268.1 to 310.9 kJ mol$^{-1}$, and the octyl chain is isolated with higher activation barrier of 377.9 kJ mol$^{-1}$. The reaction products between butyl chains and anionic fragments include trialkylphosphines, alkenes, CO and $CO_2$. In contrast, the stable octyl radical reacts with anionic fragment and forms alkyl-boranes. The calculated structural features and product distributions are in agreement with experimental data obtained from Fourier transform infrared spectroscopy, thermal gravimetric analysis (TGA) and mass spectrometry (MS) experiments.

According to quantum chemical calculations, the compact $[P_{4,4,4,8}]Cl$ IL is less stable since the activation barriers with accessible butyl chains are around 167.0 kJ mol$^{-1}$. Meanwhile, the isolated octyl chain in $[P_{4,4,4,8}]$ cation does not react with Cl anion demonstrating weaker nucleophilicity than that for [BOB] anion. These computational results are contradicted with TGA measurements where [$P_{4,4,4,8}$]Cl IL is more stable at high temperature. Although, the sample of [$P_{4,4,4,8}$][BOB] lost less weight at initial stage of test, it was shown that [BOB] anion is more stable at the initial stage of reaction. Presumably, nucleophilic anionic fragments are initially accumulated and further cause rapid thermal decomposition of [$P_{4,4,4,8}$] cation in related ILs.

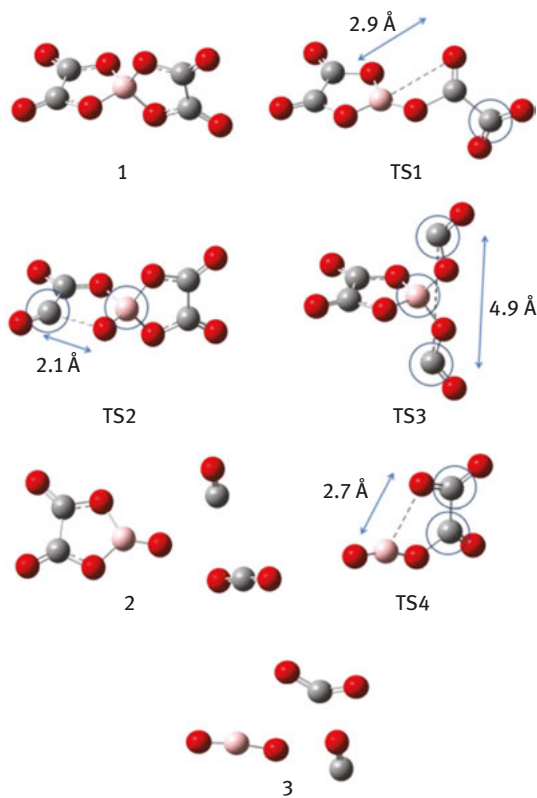

**Figure 4.4:** Optimized ionic structures of stationary points and transition states observed for thermal decomposition of an isolated [BOB] anion. The scanned coordinates in transition states are marked with dotted line. Atomic distances are presented by Å. Blue circles are attributed to imaginary frequencies: $TS_1$ (−1381.61 cm$^{-1}$), $TS_2$ (−270.30 cm$^{-1}$), $TS_3$ (−494.31 cm$^{-1}$) and $TS_4$ (−1924.23 cm$^{-1}$) as asymmetrical stretching around circled atoms.

## 4.2.4 *Ab initio* molecular dynamics simulations of ionic liquid clusters

With developed force field parameters, the structural and dynamical properties of ILs that are balanced by Coulombic and dispersive forces can be investigated on a molecular scale. For imidazolium-based ILs, the delicate intermolecular coordinations, like HB and π-type interactions that are rarely occurring simultaneously in traditional molten salts [1, 2, 10], are critical for the formation of rich ionic structures in bulk liquid and confined environment [1–3, 26, 39, 40, 49, 50, 71]. However, the nature of these interactions is distinctive and complex, and additionally, the delicate interplay among these interactions is complicated and depends on specific ion types [38–40, 51, 72, 73].

For imidazolium cations coupled with small anions, such as Cl, thiocyanate ([SCN]) and [NO₃] [38–40, 49, 51], HB and π-type interactions simultaneously occur between ionic species. Their cooperative effect promotes the formation of prominent ordered microstructures in bulk liquids. However, when imidazolium cations are associated with large anionic groups, like [NTF₂] [50, 72], both HB and π-type interactions are considerably weakened in IL phase. Large anions typically have multiple HB interaction sites and exhibit reduced HB strength and directionality in liquid environment. Additionally, these large anions take preferential *on-top* distribution above and below imidazolium rings, leading to π-type interactions being partially blocked due to anionic size effect. The delicate interplay of HB and π-type interactions among ionic species, either competitive or cooperative, becomes more complicated than this intuitive explanation implies if anions are featured with ring structures, like the chelated orthoborate families [33, 57, 70, 74–76].

We performed AIMD simulations to study the complex interplay between HB characteristics and π-type stacking features in the [MMIM][BOB] IL [77]. This IL is selected because of the relative simplicity of [MMIM] cation and [BOB] anion, but essential intermolecular interactions, like HB and π-type interactions between [MMIM] cations and [BOB] anions, are included. AIMD simulation results indicated that interactions between [MMIM] cations are stabilized by distinctive parallel π–π stacking interactions between imidazolium rings at short distance, which overtake repulsive electrostatic interactions and other weak intermolecular interactions in determining the relative distribution of neighboring imidazolium rings characterized by preferential *on-top* parallel orientations. The spatial orientations of imidazolium to neighboring oxalato rings are characterized by π–π stacking and parallel displaced offset stacking configurations at short distance and by sharp perpendicular distributions at intermediate distance, respectively. The former is stabilized by directional HB interactions between hydrogen atoms of [MMIM] cations and oxygen atoms of [BOB] anions, while the latter is dominated by attractive electrostatic interactions between ionic species. The spatial coordination pattern between intermolecular oxalato rings in [BOB] anions is balanced by repulsive Coulombic interactions and steric hindrance effect, leading to their tilted orientation in coordinating neighboring imidazolium cations in local ionic environment. Representative imidazolium-imidazolium, imidazolium-oxalato and oxalate-oxalato ionic structures are presented in Figure Figure 4.5.

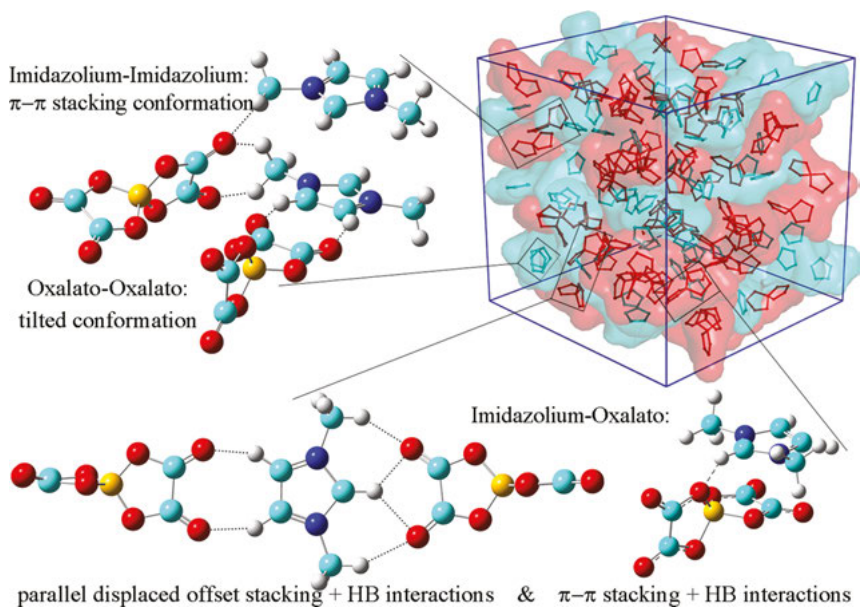

**Figure 4.5:** Representative molecular structures and orientations among close contact ionic groups. The imidazolium-imidazolium pair is featured with π–π stacking distribution. The imidazolium-oxalato pair is characterized by π–π stacking and parallel displaced offset stacking distributions, as well as HB interactions between hydrogen atoms in [MMIM] cations and oxygen atoms in [BOB] anions. The oxalato-oxalato pair is described by tilted molecular distribution that promotes HB interactions with neighboring [MMIM] cations.

# 4.3 Atomistic Molecular Dynamics Simulations

## 4.3.1 Tetraalkylphosphonium-orthoborate ionic liquid–water mixtures

For IL-related research and applications in laboratory and in industrial community, an inevitable and critical issue is the presence of impurities in IL samples [8, 59, 78–80]. As an omnipresent compound, water is one of the most common contaminators found in ILs, on one hand due to the intrinsic hygroscopic nature of some ILs, and on the other hand because many chemical processes (synthesis, extraction, etc.) involve water [33, 38, 79–86]. It has been well documented in experimental studies that even small traces of water can dramatically alter microstructures in an IL matrix, and thereafter result in significant changes in ILs' physicochemical properties, such as liquid densities [79, 83–85], diffusion coefficients [33, 82, 84–86], viscosities and rheological quantities [80–84, 86], as well as reactivity and selectivity of chemical reactions taking place in ILs [2, 3, 8, 12, 16, 18].

It was found in a recent work that the functionalization of orthoborate anions has obvious effect on residual water content in tetraalkylphosphonium-orthoborate ILs [33]. The retained water molecules are quite difficult to remove from these IL samples even after iterative purifications including vacuum drying for a few hours at 85–90 °C. The intrusion of water molecules may disturb microstructural organization in local environment, leading to a distinct change in thermodynamic properties, transport quantities and even macroscopic functional performance of ILs in practical applications. Thus, a molecular level understanding of the critical effect of residual water molecules on phase behavior of tetraalkylphosphonium-orthoborate ILs is necessary, not only because water is inevitably present in many practical applications, but also because it provides a new opportunity to tune various properties of tetraalkylphosphonium-orthoborate ILs by introducing a controllable amount of water molecules.

In order to understand why it is difficult to remove water molecules from IL matrices, we, from a thermodynamic point of view, calculated the solvation free energy of transferring one water molecule from gas phase into bulk $[P_{6,6,6,14}]$-orthoborate ILs as a function of temperature [76]. The calculated solvation free energies are positive and exhibit linear dependence on temperature, indicating that such a solvation procedure is a thermodynamically non-spontaneous process. The larger the solvation free energy the harder it is to dissolve water into IL matrix, that is, the easier it is to remove water from IL sample, which is exactly the experimental procedure to purify freshly synthesized IL sample through degassing it in a vacuum oven at elevated temperature. At given temperatures, the solvation free energy of dissolving one water molecule in four $[P_{6,6,6,14}]$-orthoborate ILs follows an order of [BMLB]>[BScB]>[BMB]>[BOB], indicating a similar possibility of removing water from IL samples. These thermodynamic simulation results are qualitatively consistent with the experimentally determined water content in four $[P_{6,6,6,14}]$-orthoborate IL samples [33]. Additionally, it takes less energy to dissolve the second water molecule within solvation shells of the first dissolved one into $[P_{6,6,6,14}]$-orthoborate IL matrices. This is attributed to an energetically favorable pairwise interaction between two water molecules, and preferential multibody interactions between water molecules and neighboring anionic groups at short distances.

The dissolution procedure does not only depend on thermodynamics but also on kinetic effects due to cooperative interactions between residual water molecules in IL matrices. In order to characterize how these dissolved water molecules behave in IL matrices, and the particular association/dissociation patterns of water molecules in coordinating neighboring $[P_{6,6,6,14}]$ cations, orthoborate anions and even water molecules nearby, we calculated the potential of mean force (PMF) between two dispersed water molecules in different $[P_{6,6,6,14}]$-orthoborate ILs as a function of water–water separation distance [76]. The water–water PMF profiles shown in Figure 4.6 indicate complex interactions of water molecules with neighboring ionic species depending on local ionic environment in IL matrices. A characteristic deep potential minimum located at around 0.28 nm is observed in four PMF profiles, indicating that intermolecular attractive interactions dominate PMF profile at short distance and favor the formation of a close contact water dimer complex through HB interactions. Additionally, this water dimer complex strongly coordinates with neighboring anionic species and forms stable ring structures through HB interactions. It takes approximately 28 kJ/mol in [$P_{6,6,6,14}$][BOB], [$P_{6,6,6,14}$][BMLB] and [$P_{6,6,6,14}$][BMB] ILs, and around 38 kJ/mol in [$P_{6,6,6,14}$][BScB] IL, respectively, to break these ring structures before pulling water molecules to larger distances. As separation distance between two water molecules slightly increases, there is still not enough space between water molecules for one ion to squeeze in due to its relatively large ionic size. This leads to unstable intermediates contributing to gradually increased water–water intermolecular potential energies until the formation of an ion-separated metastable water association structure at a larger distance.

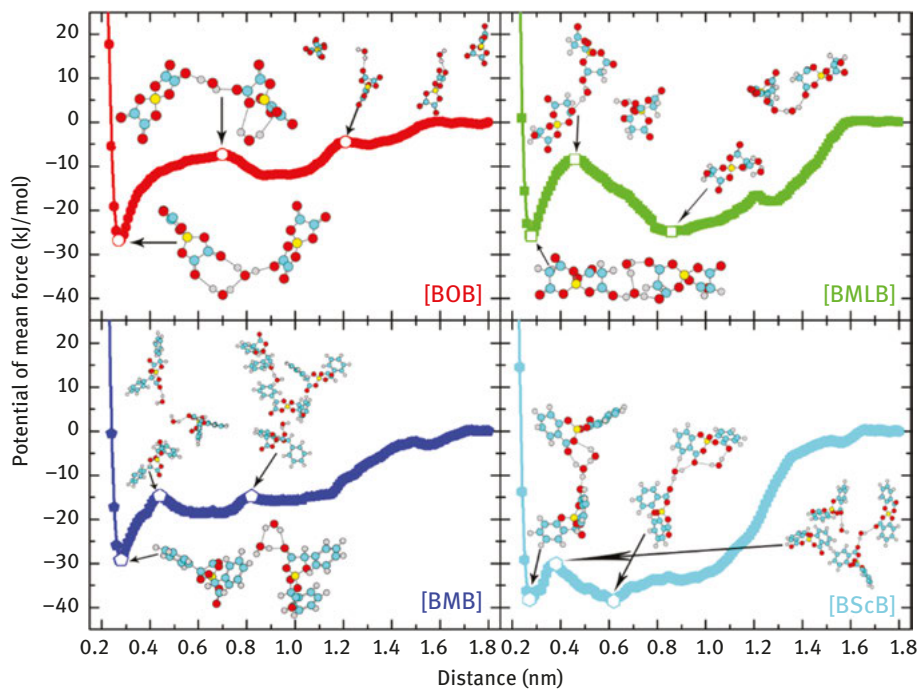

**Figure 4.6:** Potential of mean force results between two dispersed water molecules, and three key intermediate configurations of dispersed water molecules in coordinating neighboring orthoborate anions in four [$P_{6,6,6,14}$]-orthoborate IL matrices obtained from atomistic simulations at 333 K.

The water–water association/dissociation patterns at larger separation distances are characterized by particular features depending on intrinsic molecular structures of orthoborate anions and delicate interactions of dispersed water molecules with surrounding ionic species. On the one hand, the hydrogen atoms in water molecules are exclusively coordinated with oxygen atoms in hydrophilic C=O groups in ortho-borate anions. The multiple hydrogen bonding acceptor sites (oxygen atoms) in orthoborate anions blur some insignificant metastable ion-separated potential mini-mum in corresponding PMF profiles. On the other hand, the introduction of either small methylene groups or large aromatic rings to [BOB] anion leads to distinct spatial coordination pattern of dispersed water molecules around neighboring ortho-borate anions, as shown in Figure 4.7. In [BOB] and [BMLB] systems, the dispersed water molecules are preferentially coordinated with anions in equatorial region of oxalato ring planes due to directional HB interactions. However, in [BMB] and [BScB] ILs, water molecules are specifically localized around central polar segments to avoid direct contact with hydrophobic phenyl rings in [BMB] and [BScB] anions. Such a distinct spatial distribution of water molecules embedded in cavities between neigh-boring ionic species contributes to the formation of intermolecular hydrogen bonds between water molecules and anionic groups, and further mediates local ionic structures through ion    water    ion multiple complexes. By pulling two water mole-cules further away, the PMF oscillation becomes weak, indicating that interactions between dispersed water molecules are partially or totally screened by intervening multiple ionic species in between.

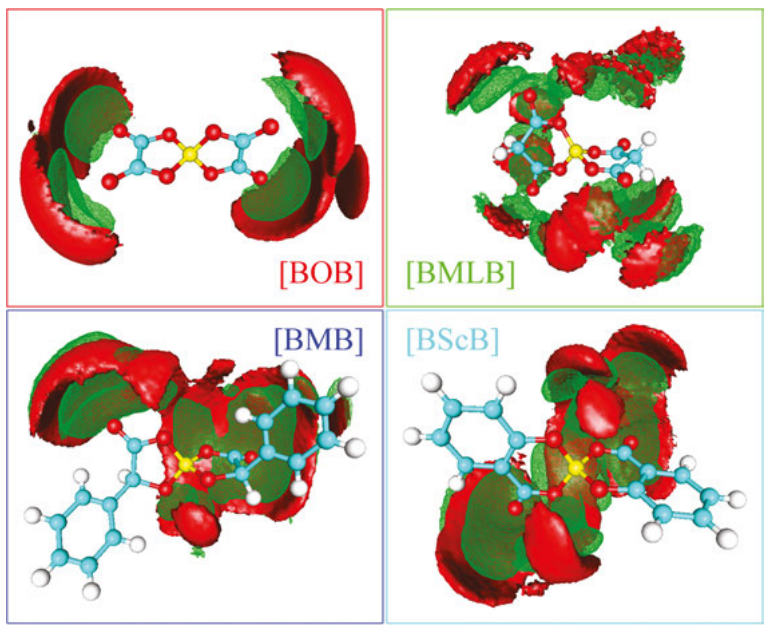

**Figure 4.7:** Spatial probability distribution functions of oxygen (solid red surface) and hydrogen (meshed green surface) atoms in water molecules around orthoborate anions in the first solvation shell obtained from atomistic simulations at 333 K. The red and green contour surfaces are drawn at 4.0 times of the average number density of the corresponding atoms in bulk system.

Comparing the potential depth of metastable ion-separated water association structure and the corresponding water–water separation distance in respective PMF profile, as well as the activation energy barrier from stable water dimer complex to metastable ion-separated water association structure, it can be further identified that the introduction of hydrophobic groups, either the small methylene unit or the large aromatic moiety, can essentially decrease the potential depth of ion-separated water association structure and the corresponding activation energy barrier, as well as shift the latter configurations to short separation distances. Integrating the activation energy barriers, solvation free energies and the residual water contents in four $[P_{6,6,6,14}]$-orthoborate ILs, we proposed that the removal of water molecules from IL matrices follows a two-step procedure. The separation of coupled water molecules within a heterogeneous IL matrix is the primary pathway, and the subsequent removal of water from bulk ILs is the secondary procedure in purification process. From a structural point of view, the introduction of hydrophobic units into central polar segments and the resulting conjugated structures in orthoborate anions can decrease the activation energy barrier for the separation of bounded water molecules, and hence can effectively reduce water content in the corresponding IL samples. This provides a valuable guidance for future design and synthesis of new orthoborate anions with different chemical polarities and possibly varied functional performance in practical applications.

In these four $[P_{6,6,6,14}]$-orthoborate ILs, the $[P_{6,6,6,14}][BOB]$ IL holds special significance. It was found that the residual water content in freshly synthesized $[P_{6,6,6,14}]$ [BOB] IL sample is 2.3–2.5 wt% [33], which corresponds to a water mole fraction of approximately 0.5 in this sample. That is, the ratio of $[P_{6,6,6,14}][BOB]$ ion pairs and water molecules is 1:1, which might suggest interesting ionic structures like water molecules residing between close contact ion pair structures. By performing extensive atomistic simulations, four distinct compositional regimes were identified concerning the evolution of microscopic liquid organization, local ionic structures, volumetric quantities and translational and rotational mobilities of ionic species in $[P_{6,6,6,14}][BOB]$ IL–water mixtures as water concentration increases [74, 75]. The variations of liquid densities and translational diffusion coefficients of ionic species in $[P_{6,6,6,14}][BOB]$ IL–water mixtures as water concentration increases are illustrated in Figure 4.8.

–   In neat ILs and IL–water mixtures with water mole fractions $x_{water} \leq 0.5$, the $[P_{6,6,6,14}]$ cations and [BOB] anions are closely coupled together via strong electrostatic interactions. The microscopic liquid organization is characterized by a connected apolar network consisting of volume-occupying alkyl substituents in $[P_{6,6,6,14}]$ cations and isolated polar domains composed of central polar segments of $[P_{6,6,6,14}]$ cations, [BOB] anions and water molecules. Such a heterogeneous local ionic environment leads to an exponential increase in translational diffusion of ionic species in $[P_{6,6,6,14}][BOB]$ IL–water mixtures as water concentration increases. Most of the added water molecules are dispersed and preferentially associated with neighboring ionic species, leading to the local ionic environment characterized by solvent-shared ion pairs through cation–water–anion triple complexes. The restricted distribution of water molecules in local ionic environment results in a constrained reorientation of water molecules on a cone surface due to their dual nature in coordinating neighboring ionic species through HB interactions.

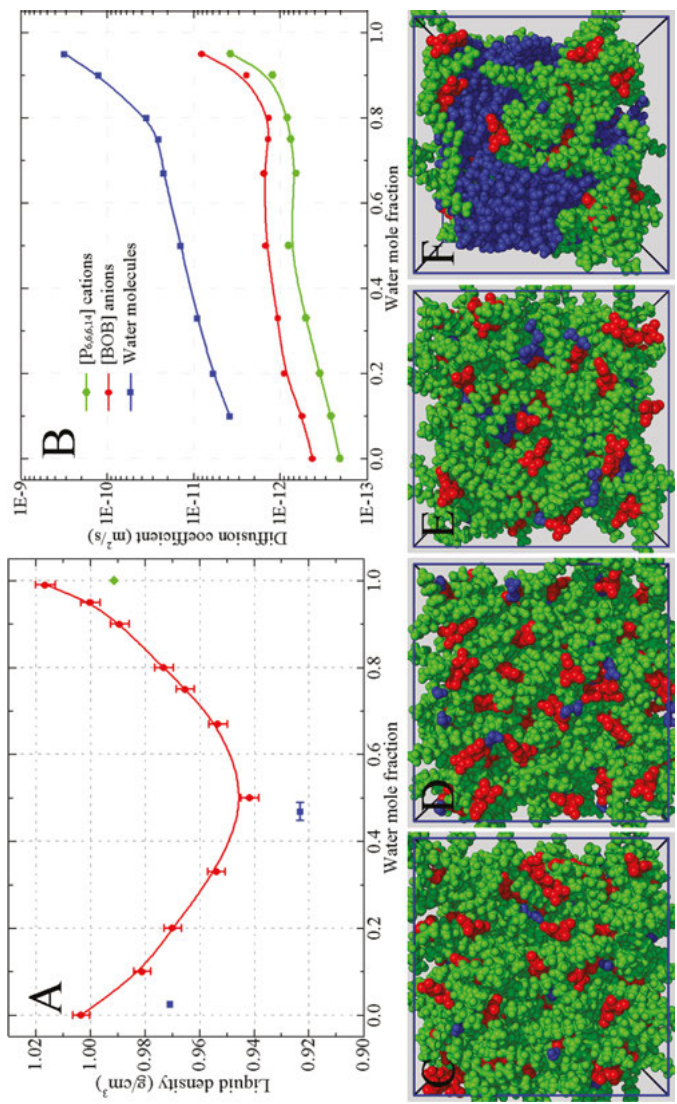

**Figure 4.8:** (A) Liquid densities of [P$_{6,6,6,14}$][BOB] IL water mixtures with different water mole fractions at 333 K obtained from atomistic simulations (solid red circles) and determined from experimental measurements (solid blue squares) of two representative [P$_{6,6,6,14}$][BOB] IL samples with residual water contents of 0.06 and 2.30–2.50 wt%, respectively. The solid green diamond corresponds to an experimental density of neat water at 333 K. (B) Translational diffusion coefficients of [P$_{6,6,6,14}$] cations, [BOB] anions and water molecules in [P$_{6,6,6,14}$][BOB] IL–water mixtures with different water mole fractions at 333 K obtained from atomistic simulations. Representative snapshots of [P$_{6,6,6,14}$][BOB] IL–water mixtures with water mole fractions of x$_{water}$ = 0.33 (C), 0.5 (D), 0.8 (E), and 0.95 (F). The [P$_{6,6,6,14}$] cations, [BOB] anions and water molecules in these mixtures are represented by green, red and blue beads, respectively.

– In IL–water mixtures with intermediate water mole fractions of $0.5 < x_{water} \leq 0.8$, large water clusters appear and dominate size distribution of water aggregates since there is no sufficient void space to accommodate more water molecules. A distinct chain-like aggregate characterized by anion$\cdots(H_2O)_n\cdots$anion structure serves as bridge connecting more anionic species between isolated polar domains, as well as mediating their relative distribution and orientation in IL–water mixtures. The local ionic organization of these mixtures is characterized by solvent-mediated ion pairs. This leads to enhanced spatial correlations between ionic species, and thus considerably slows down translational and orientational mobilities of ionic species in IL–water mixtures.

– In IL–water mixtures with water mole fractions of $0.8 < x_{water} \leq 0.95$, water molecules are dynamically percolated throughout the entire simulation box and constructing a water network, leading to microscopic liquid environment described by interpenetrating polar and apolar networks. The percolation of polar domains within apolar framework promotes a rapid increase in translational diffusion of ionic species in these water-concentrated mixtures. Albeit there is a large amount of water in these mixtures, the central polar segments of $[P_{6,6,6,14}]$ cations and [BOB] anions remain in close proximity to each other through sharing one or more water molecules. This complex structure is rationalized by strong electrostatic interactions and favorable HB interactions between ionic species.

– In water concentrated mixtures with water mole fractions of $x_{water} > 0.95$, a further progressive dilution of IL–water mixtures leads to a percolation limit of IL in water, that is, upon further dilution, the connected apolar network (already stretched to its limit) starts to break up and loses its continuous nature. The local ionic environment is characterized by loose micelle-like aggregates in a highly branched water network.

The striking evolution of microstructures, local ionic environment, translational and orientational diffusion of ionic species in $[P_{6,6,6,14}]$[BOB] IL–water mixtures indicate that these mixtures are characterized by particular microstructural and dynamical heterogeneities. Such a spatio-temporal heterogeneity can be attributed to a competition between favorable HB interactions and strong Coulombic interactions between central polar segments in ionic species, and persistent dispersion interactions between hydrophobic alkyl chains in $[P_{6,6,6,14}]$ cations.

## 4.3.2 Neat ionic liquids in confined environment

### 4.3.2.1 [BMIM][PF$_6$] on neutral graphene surface

Either ILs used as lubricants between sliding contacts in tribology [12, 14, 15, 32, 33], as electrolytes in electrochemical energy devices [1, 7, 12, 16, 17, 21, 29, 31, 71, 87] or as liquid absorbents for $CO_2$ capture from fossil-fuel burning power plants [12, 13, 19, 20, 28], a common feature is the presence of interfaces between ILs and solid, gas phases [1, 7, 12–17, 19, 20, 26, 28–30, 32, 71, 87–93]. Experimental investigations of ILs under confinement indicate that physicochemical properties (melting points, surface tensions, chemical reactivities, etc.) and microstructures of ILs in interfacial regions are different to those in bulk phase [1, 7, 12, 14, 15, 17, 19, 23, 26, 28–30, 32, 71, 87–89, 93]. The direct recoil spectrometry measurements indicated a perpendicular distribution of imidazolium ring planes and a parallel orientation of butyl chains in [BMIM] cations in IL–gas interfacial region [94]. However, experimental characterizations based on sum frequency generation (SFG), high-resolution Rutherford backscattering spectroscopy and X-ray reflectivity observations [95–98] suggested that imidazolium rings prefer to take flat orientations along IL–gas interface, whilst butyl chains are loosely packed together and protruded from liquid phase into gas phase with a certain tilt. Additionally, in IL–solid interfacial region, nuclear magnetic resonance experiments showed that interfacial ionic structures of supported [BMIM][PF$_6$] IL depend strongly on solid surface types and the thickness of the confined IL film [93].

For absorbed [BMIM][PF$_6$] IL on neutral graphene surface with finite film thickness, it was observed in atomistic simulations that the confined IL film thickness has a significant effect on interfacial ionic structures and ordering orientational distributions of [BMIM][PF$_6$] ion pairs in IL–graphene and IL–gas interfaces [99]. Interfacial monolayers are formed in IL–gas interfacial region with ordering ionic structures. With an increase in IL film thickness, the orientations of [BMIM] cations in interfacial monolayers change gradually from dominant flat distributions along IL–gas interface to that characterized with several favorable orientations with different proportions, as shown in Figure 4.9. In these favorable orientations, the main distribution of imidazolium rings is their parallel distributions along IL–gas interface beneath the exposed outermost layer and perpendicular orientations with tilted angles toward IL–gas interface at the exposed outermost layer. The outmost layer is populated with alkyl groups and is imparted with hydrophobic feature. Distinct ionic layers with well-regulated interpenetrating polar and nonpolar networks are formed in the vicinity of IL–graphene interfacial region. The imidazolium rings lie preferentially flat on IL–graphene interfacial region, with the methyl and butyl chains in [BMIM] cations stretched out along graphene surface.

The presence of solid surface and formation of dense interfacial layers further complicate ILs' dynamical properties compared with those in bulk regions. It is shown that the dynamical quantities of the confined [BMIM][PF$_6$] ionic groups are highly heterogeneous depending on their relative positions in confined IL film, as characterized by mean square displacements of specific groups [100] shown in Figure 4.10. In IL–gas interfacial region, the relaxation and diffusion of terminal carbon atoms in butyl chains of [BMIM] cations are much faster than that in other layers of confined IL film and that in bulk region of simulation system without confinement, due to their liberated motion in this interfacial region. In IL–graphene interface region, the dynamical heterogeneities of ionic groups are embodied both in their overall diffusions and in the diffusive components in parallel plane along solid surface and in perpendicular direction to solid surface, respectively. The overall translational mobility of [BMIM][PF$_6$] ionic species is characterized with slaved diffusion, and it takes much longer time for them to leave the cage formed by surrounding counterions before reaching the true diffusive regime than that without confinement.

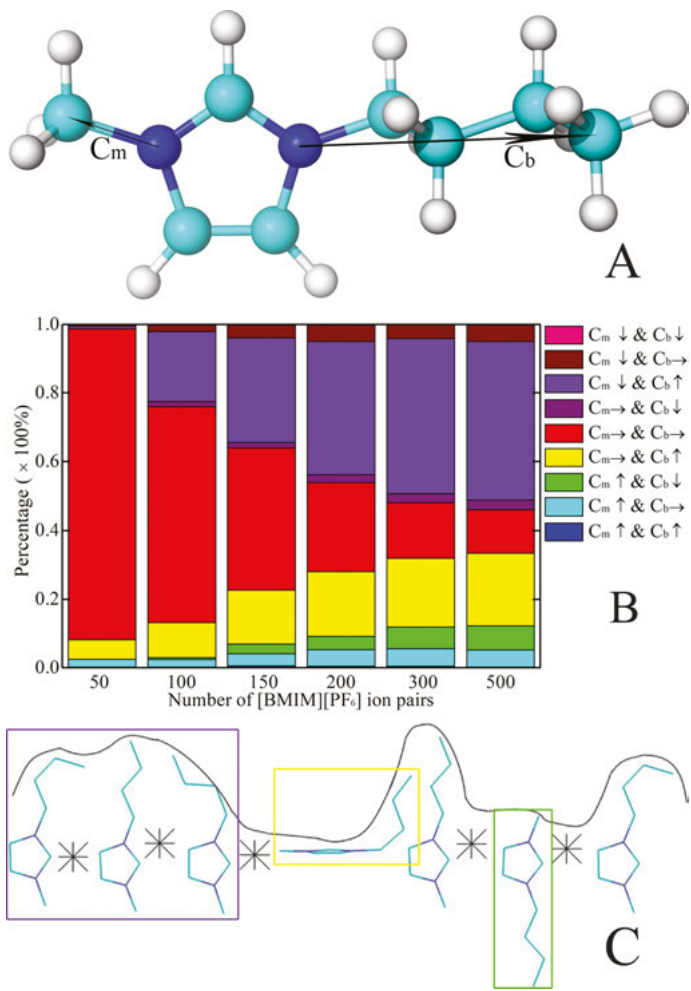

**Figure 4.9:** (A) Definition of $C_m$ and $C_b$ vectors in [BMIM] cationic framework and (B) probability distributions of these two vectors in the same ion for [BMIM] cations in IL–gas interfacial region in simulation systems consisting of varied numbers of [BMIM][PF$_6$] ion pairs on graphene surface. The notations →, ↓ and ↑ indicate that $C_m$ or $C_b$ vectors being parallel or perpendicular to the IL–gas interface with terminal carbon atoms projected into bulk region or protruded into gas phase, respectively. (C) Representative configurations of [BMIM] cations in IL–gas interfacial region.

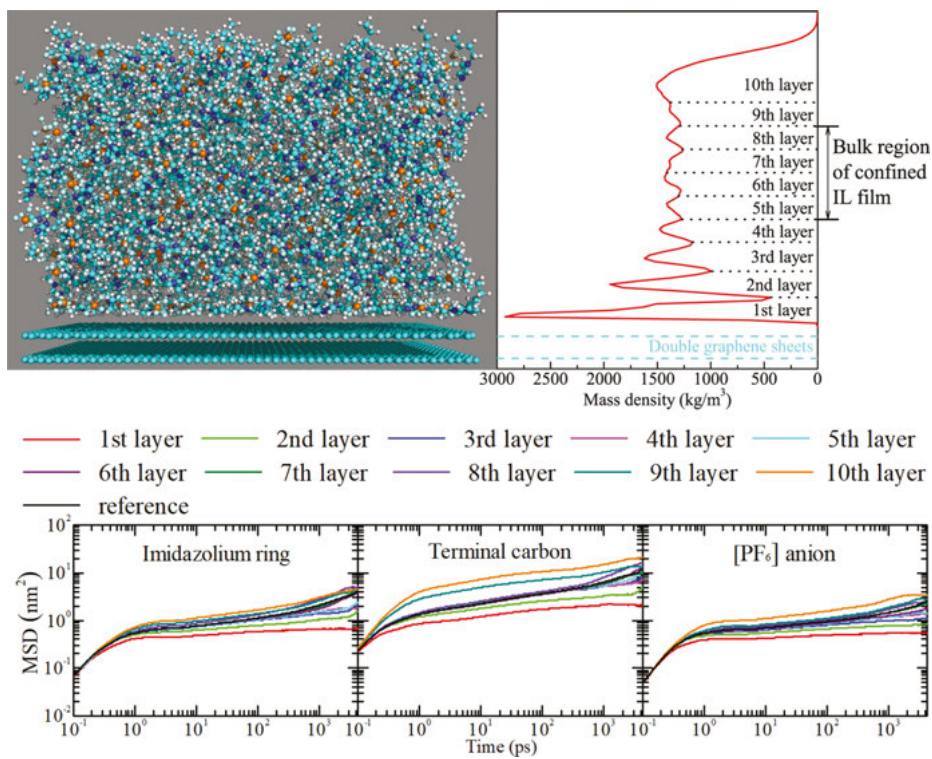

**Figure 4.10:** Representative snapshot of confined IL film consisting of 500 [BMIM][PF$_6$] ion pairs on a neutral graphene surface and the division of this IL film into ten layers based on mass density profile. Mean square displacements of imidazolium rings and terminal carbon atoms of butyl chains in [BMIM] cations and of [PF$_6$] anions in different layers of confined IL film, as well as that in simulation system without confinement.

The particular dynamical heterogeneity of confined [BMIM][PF$_6$] ionic groups is intrinsically related to their orientational preference and microscopic ionic structures in IL–graphene and IL–gas interfacial regions. The confinement effect induced by neutral graphene surface and formation of two-dimensional interpenetrating mesophase contribute to the slaved diffusion of [BMIM][PF$_6$] ion pairs in IL–graphene interfacial region. However, such a confinement effect is short-ranged, and its influence on dynamical properties is limited in bottom layers of confined IL film and becomes negligible with the increase of film thickness. For IL film with enough thickness, such as larger than 3 nm, the exposed outmost layer is populated with alkyl groups, mainly terminal carbon atoms of butyl chains of [BMIM] cations, which facilitates their liberated motion and hence contributes to their fast diffusion in IL–gas interfacial region. Beneath such exposed layer in IL–gas interface, the microscopic ionic structure resembles that in bulk region of confined IL film, and hence exhibits similar dynamical properties. It is the spatial structural heterogeneity of confined ion pairs in interfacial region that contributes directly to the striking dynamical heterogeneity. Both spatial and temporal heterogeneities of confined [BMIM][PF$_6$] ion pairs experiencing in interfacial regions are important for understanding the interfacial phenomena occurring in IL–solid and IL–gas interfacial regions before advancing their applications in related areas.

## 4.3.2.2 [BMIM]-based ionic liquids on charged quartz surfaces

The confined ionic species, however, exhibit distinct microstructures and ordering orientations in IL–solid interfacial region if solid surfaces are characterized by positive or negative charges [1, 7, 17, 30, 71, 87]. On charged quartz surfaces, the atomic force microscope (AFM) measurements revealed that the formation of interfacial layers of solvated ILs is strongly related to quartz surface charge and roughness [1, 26, 101]. The SFG spectroscopic signature indicated that imidazolium rings prefer to lie on quartz surface with the attached alkyl chains taking on tilted orientations [89, 90, 92]. However, this qualitative conclusion was summarized based on a simple assumption that each confined cation behaves as a rigid entity, and on limited orientational information of "reporter" groups, such as the vibrational frequency of C—H bonds on imidazolium rings.

Due to the lack of specific spectroscopic signatures for aliphatic chains, the distribution and orientation of alkyl chains in [BMIM] cations are not directly investigated. Therefore, we designed two chemically different quartz surface models to mimic the adsorption of [BMIM]-based ILs on synthesized and catalytic quartz surfaces [90–92, 102]. The dangling silicon atoms in one bare quartz surface are saturated with silanol Si(OH)$_2$ groups, whereas those in the other one are fully hydrogenated and covered by silane SiH$_2$ groups, respectively [103], as shown in Figure 4.11. Such an intrinsic difference in local chemical composition leads to the quartz surfaces saturated with Si(OH)$_2$ and SiH$_2$ groups characterized by negative and positive charges, respectively, which result in distinct stacking behavior of absorbed [BMIM]-based ILs on these two quartz surfaces.

Atomistic simulation results revealed that dense ionic layers, characterized by distinct mass, number, charge and electron densities, are formed in quartz interfacial region. The orientational preferences of confined ionic groups are characterized with different features depending on quartz surface charges, ionic sizes and molecular geometries of anionic groups. In positively charged $SiH_2$ interfacial region, the anionic groups are particularly absorbed on solid surface due to strong electrostatic interactions. The main axes of asymmetric [TFO] and [NTF$_2$] anions are perpendicular and parallel to $SiH_2$ surface, respectively. However, an opposite effect is observed in negatively charged $Si(OH)_2$ interfacial region. The [BMIM] cations, either coupled with spherical ([BF$_4$] and [PF$_6$]) or aspherical ([TFO] and [NTF$_2$]) anions, are exclusively absorbed onto negatively charged $Si(OH)_2$ surface. The imidazolium rings lie predominantly perpendicular to $Si(OH)_2$ surface, with the corresponding methyl and butyl chains elongated along $Si(OH)_2$ surface, respectively. The anions exhibit random orientations in subsequent anionic layer, due to the partially screened intermolecular interactions between anions and atoms on $Si(OH)_2$ surface. Typical configurations of absorbed [BMIM] cations and their coupled [BF$_4$], [PF$_6$], [TFO] and [NTF$_2$] anions in $Si(OH)_2$ and $SiH_2$ interfacial regions are shown in Figure 4.11.

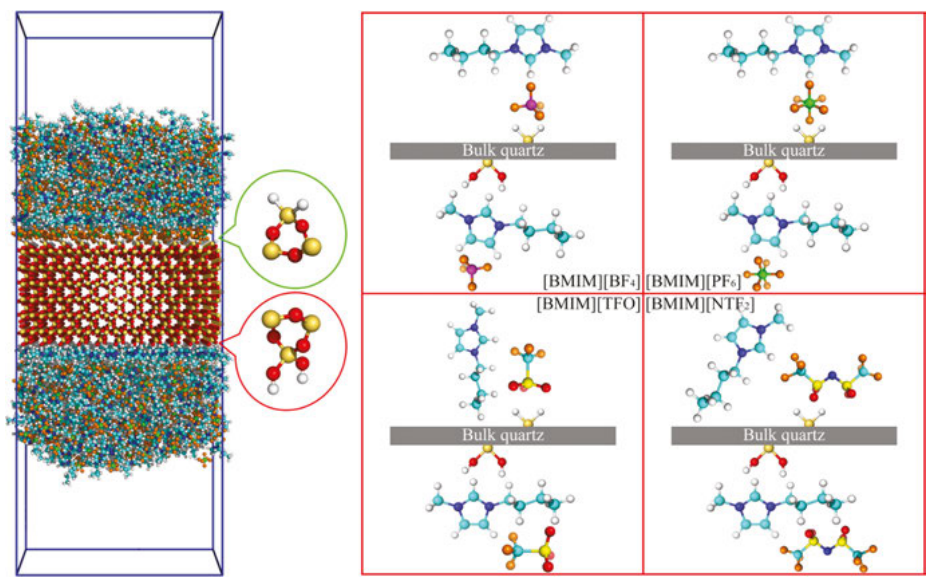

**Figure 4.11:** Representative configurations of [BMIM] based ILs confined in charged $SiH_2$ (top) and $Si(OH)_2$ (bottom) quartz interfacial region.

The distinct structural and orientational preferences of confined [BMIM] cations, and their coupled anions in IL–quartz interfacial regions are intrinsically related to quartz surface charge densities, molecular ionic sizes and geometries, as well as preferential intermolecular interactions between ionic groups and atoms constituting quartz solid surfaces. Given that ILs serve as both charged species and solvent molecules in confined environment, large ion concentration corresponds to small Debye length, indicating that electrostatic interactions are partially screened, and thus other factors, like crowding effect originated from solid surface templating issue, become important. The subtle interplay between screening and crowding effects contributes to the formation of electric double-layer structures in charged interfacial region, which may provide an opportunity to unveil intrinsic ionic structures in controlling macroscopic functional performance of electric supercapacitors [7, 16, 20, 29, 30, 71, 87].

## 4.3.2.3 [P$_{6,6,6,14}$][BMB] ionic liquid on charged gold electrodes

For all the confined [BMIM]-based ILs in solid interfacial regions, either on neutral graphene surface or on charged quartz surfaces, the charge density on solid surfaces is constant. In these confined systems, it is straightforward to study interfacial structures of confined ionic species on solid surfaces with specific charge densities, but not convenient to investigate the dependence of microstructural ordering distributions and orientations of confined ionic species on surface charge densities in IL–solid interfacial region. The application of external electric fields on solid surfaces with different interfacial charge densities, as verified in AFM and surface force apparatus (SFA) experimental characterizations [1, 26, 71, 104–107], is an effective way to induce molecular position fluctuations in confined environment, leading to the accumulation of specific ionic species in interfacial region and finally resulting in reconstruction of interfacial layering structures.

In order to understand the dependence of interfacial structures of confined ionic species on solid surface charge densities, we performed intensive atomistic simulations to probe interfacial ionic structures and molecular arrangements of [P$_{6,6,6,14}$] [BMB] IL (chosen owing to its excellent tribochemical properties in mechanical engineering contacts [33, 108]) confined between neutral and charged gold electrodes with controllable surface charge densities [109]. Detailed analyses of simulation results indicate that the interfacial chemical compositions, molecular arrangements of [P$_{6,6,6,14}$][BMB] IL are different depending on surface charge densities to gold electrodes. For [P$_{6,6,6,14}$][BMB] IL confined between neutral electrodes, due to the interfacial layer and a subsequent intermediate layer are formed before reaching bulk region of confined IL film. The innermost layer consists of both [P$_{6,6,6,14}$] cations and [BMB] anions, which take compact ionic structures and checkerboard molecular arrangement in interfacial region. In this mixed innermost layer, both hexyl and tetradecyl chains in [P$_{6,6,6,14}$] cations lie preferentially parallel along electrodes, and the most probable configuration of oxalato and phenyl rings in [BMB] anions is characterized by consecutive parallel and perpendicular arrangement adjacent to neutral electrodes, respectively.

As gold electrodes get electrified but with low surface charge densities (<20 μC/cm$^2$), the mixed innermost layer thickness gradually increases as that in surface charge density, due to a gradual accumulation of [P$_{6,6,6,14}$] cations and [BMB] anions, and their counterions being squeezed out from the innermost layer adjacent to negatively and positively charged electrodes, respectively. The effect of charging electrodes has little influence on the molecular alignment of hexyl and tetradecyl chains in $[P_{6,6,6,14}]$ cations along negatively charged electrodes due to their delocalized charge distribution within cationic framework and their saturated distribution in the innermost cationic layer. However, charging gold electrodes results in new orientational patterns for oxalato rings in the same [BMB] anions from parallel–perpendicular orientation to that characterized by constraint molecular arrangement with a tilted angle of 45° from positively charged electrode. In the meantime, the molecular distributions of phenyl rings are alerted accordingly due to their direct bonding to oxalato rings through flexible C—C bonds in [BMB] anions. Upon further charging gold electrodes with surface charge densities equal or higher than 20 $\mu C/cm^2$, distinctive innermost interfacial layers exclusively consisting of $[P_{6,6,6,14}]$ cations and [BMB] anions are formed adjacent to negatively and positively charged electrodes, respectively. This implies a templating effect in producing compact and tightly bounded innermost layers closest to charged electrodes as surface charge density increases. Such an interfacial effect will in turn alter packing ionic structures in subsequent layers, and so forth, resulting in enhanced compact interfacial structures in confined environment. It is expected that more energies are needed for a probe to rupture, to penetrate and to displace the innermost ionic layer due to the absorbed ionic species being strongly bounded to oppositely charge electrodes, as indicated from previous AFM and SFA measurements [104–107]. The orientations of oxalato and phenyl rings in [BMB] anions are described by broad and featureless characteristics before distinctive coordination pattern observed for [BMB] anions adjacent to positively charged electrodes with surface charge density of 100 $\mu C/cm^2$.

It is noteworthy that $[P_{6,6,6,14}]$ cations and [BMB] anions exhibit different responses to changes in external electric field generated from charged electrodes, which will have a profound effect on interfacial friction when this IL is used to lubricate gold engineering surfaces in technical applications. These atomistic simulation results are helpful in elucidating interfacial changes in ionic structures and molecular arrangements of $[P_{6,6,6,14}]$[BMB] IL confined between gold electrodes, and may provide more physical insight in further investigation of ILs' electrotunable friction response and lubrication mechanism in mechanical engineering systems.

## 4.3.3 Simulation algorithms for calculation of ionic liquid viscosity

The most immediate way to calculate the liquid viscosity is to perform a shear flow simulation. This can be done by using SLLOD equations of motion [110]. They have been used since 1980s to calculate the liquid viscosities of alkanes, Lennard-Jones liquids and liquid crystals [111–113]. When these equations are used, a planar Couette velocity field, $\mathbf{u} = \dot{\gamma}z\mathbf{e}_x$, that is, a velocity in the $x$-direction varying linearly in the $z$-direction, is added to the ordinary Hamiltonian equations of motion

$$\dot{q}_{i\alpha} = \frac{p_{i\alpha}}{m_\alpha} + \dot{\gamma}e_x z_{i\alpha} \tag{4.1a}$$

and

$$\mathbf{p}_{i\alpha} = \mathbf{F}_{i\alpha} - \dot{\gamma}\mathbf{e}_x p_{i\alpha z} - \varsigma\mathbf{p}_{i\alpha} \tag{4.1b}$$

where $\mathbf{p}_{i\alpha}$ and $\mathbf{q}_{i\alpha}$ are the linear momentum relative to streaming velocity, the position of atom $\alpha$ of molecule $i$, $\dot{\gamma}$ is the velocity gradient or shear rate, $m_\alpha$ is the mass of atom $\alpha$ and $\zeta$ is a thermostatting multiplier removing the excess heat generated by the shear field. The algebraic expressions for this multiplier can be selected in a few different ways depending on the desired ensemble. The edge effect vanishes by applying Lees–Edwards sliding brick boundary conditions. The viscosity $\eta$, in the Newtonian or linear regime is obtained as the zero field limit of ratio of the $zx$-component of shear stress, $\sigma_{zx}^s$, and the velocity gradient

$$\eta = \lim_{t \to \infty} \sigma_{zx}^s(t)/\dot{\gamma}. \tag{4.2}$$

According to linear response theory, the zero shear rate limit of the viscosity thus obtained should be exactly the same as the one given by the Green–Kubo relation for the viscosity

$$\eta = \frac{V}{10k_B T} \int_0^\infty dt < \bar{\sigma}^s(t) : \bar{\sigma}^s(0) >_{eq}, \tag{4.3}$$

where $V$ is the volume of the system, $k_B$ is Boltzmann constant, $T$ is the absolute temperature, $\bar{\sigma}^s = \frac{1}{2}\left(\bar{\sigma} + \bar{\sigma}^T\right) - \frac{1}{3}Tr(\bar{\sigma})$ is the symmetric traceless part of stress tensor and the subscript "$eq$" denotes that the average is evaluated in an equilibrium ensemble. Since shear flow simulations and the Green–Kubo relation should give the same results, they provide an important cross-check of atomistic calculations, that is, the shear rates in eq. (2) fall into a linear regime and that the correlation functions in eq. (3) have decayed. It is also a consistency test of the computer program used to perform simulations.

The shear flow algorithm and equilibrium Green–Kubo method have recently been adopted to calculate the liquid viscosities of [P$_{6,6,6,14}$][BMB] and [P$_{6,6,6,14}$][BOB] ILs. The selection of these two IL systems is due to their potential technological applications such as lubricants and electrolytes for fuel cells and others. It should be noted that it is still a challenge in a proper sense of word to calculate liquid viscosities of [$P_{6,6,6,14}$][BMB] and [$P_{6,6,6,14}$][BOB] ILs owing to their complex molecular structures. Since these ionic molecules are rather large, there are polarization and charge transfer phenomena that can be explicitly handled in classical molecular dynamics simulations [114, 115]. However, this is also time consuming in practice for the abovementioned tetraalkylphosphonium-orthoborate ILs, because very long simulations are needed to evaluate their liquid viscosities. Therefore, we adopted a feasible procedure by using charge scaling factors between ionic species to calculate the liquid viscosities of these ILs.

The preliminary results of atomistic simulations on these ILs have shown that the viscosities of [$P_{6,6,6,14}$][BMB] IL system do not change within statistical uncertainty when system size is increased from 96 to 735 ion pairs [116]. This is advantageous when the Green–Kubo relation is evaluated since thermal fluctuations are larger in smaller systems, which means that the time correlation functions converge faster. It has also been established that the viscosities obtained using the Green–Kubo method from equilibrium atomistic simulations agree with those obtained from shear flow simulations in the linear regime, which is in accordance with linear response theory. This provides a significant consistency test of simulation algorithm and shows that the computation time is long enough to obtain reliable values for the viscosity. As an example of computational results, we show the viscosities of an equimolar mixture of [$P_{6,6,6,14}$][BMB] and [$P_{6,6,6,14}$][BOB] ILs as a function of temperature between 363 and 423 K in Figure 4.12. These viscosities have been evaluated using Green–Kubo relation (eq. (3)) with a charge scaling factor of 0.8. It is obviously shown that the liquid viscosity of this mixture can be described by a rather simple Arrhenius expression over the studied temperature range, despite the complexity of this simulation system consisting of [$P_{6,6,6,14}$] cations and [BMB] and [BOB] anions.

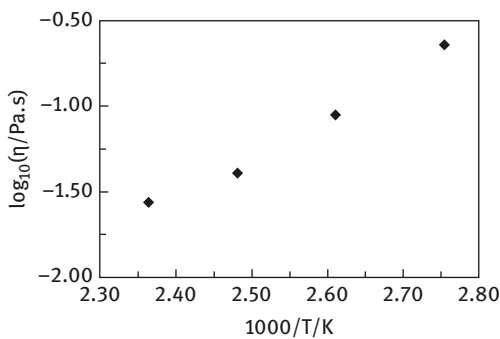

**Figure 4.12:** Liquid viscosities of an equimolar mixture of [$P_{6,6,6,14}$][BMB] and [$P_{6,6,6,14}$][BOB] ILs at different temperatures evaluated using Green–Kubo relations with a charge scaling factor of 0.8 for ionic species.

### 4.3.4 Atomistic simulations and small-angle X-ray scattering studies of alkylammonium nitrate with cosolvents

EAN was studied already in the year of 1914 by Paul Walden in a pioneering study, which is considered to mark the beginning of investigation on ILs [4, 12]. After more than 100 years, many features of EAN, and more in general of alkyl ammonium nitrate, and their mixtures with other solvents, are still attracting considerable interest.

We have recently investigated both experimentally and by means of molecular modeling the mixtures of this class of IL with important cosolvents such as N-methyl-2-pyrrolidone (NMP) and acetonitrile (ACN) [79, 81, 117, 118]. IL mixtures with cosolvents are macroscopically homogeneous but are typically characterized by microheterogeneity at nanoscale level. Small-angle X-ray scattering experiments (SAXS) on some IL mixtures with some cosolvents have revealed a pronounced scattered intensity in low $q$ region of SAXS pattern [119–121]. Such phenomenon, termed as low $q$ excess (LqE), is higher at low concentration of IL. In a recent investigation, some of us found that the mixtures of EAN with ACN display the same phenomenon. Furthermore, the LqE phenomenon observed at low concentration of EAN is the highest reported to date [118].

A combined experimental (volumetric measures, wide and small-angle X-ray scattering) and computational study performed on the whole concentration range of this EAN–ACN mixture allows to understand the microscopic origin of this finding. The partial molar volumes suggest that addition of small amount of EAN to ACN does not lead to a homogenous solution at molecular level, and that EAN molecules self-associate. The isobaric thermal expansivity data pointed to a similar conclusion concerning the lack of homogeneity at molecular scale. The nonhomogeneity of local structure leads to density fluctuations, and this is the origin of LqE observed at EAN molar fraction ($x_{EAN}$) lower than 0.5. Through the calculation of entity of density fluctuation at different EAN molar fractions, it can clearly be seen that the highest fluctuations occur at the lowest EAN concentration and that they decrease exponentially with increasing EAN molar fraction.

Atomistic simulations of these mixtures in the whole concentration range allowed reaching an understanding of these EAN–ACN systems at molecular level. The general Amber force field was used for bonded and vdW parameters [122], while partial charges were calculated at B3LYP/aug-cc-pVTZ level of theory, using a restricted electrostatic potential procedure. The initial simulation boxes for atomistic simulations were prepared using the same molar fraction as that in experimental samples. A dielectric constant of 1.8 was used in simulation with EAN, which is equivalent to charge scaling of 0.75.

Analysis of atomistic simulation trajectories at $x_{EAN} = 0.1$ reveals that EAN molecules organize in wormlike structure surrounded by ACN molecules. These molecular structures are constituted mainly by EAN, and their density is higher than that of neat IL due to a solvophobic effect and is much denser than ACN. Similar organization, although less pronounced, is observed for $x_{EAN}$ molar fraction of 0.3 and 0.5, while at higher EAN content the mixtures are homogeneous at molecular level, with ACN molecules dissolved in EAN matrix. The structural heterogeneity is observed in atomistic simulations only for mixtures that display LqE in SAXS patterns. This clearly indicates that the physical objects generating density variations that induce LqE are wormlike structures formed mainly by EAN ionic species.

# 4.4 Coarse-Grained molecular dynamics simulations

## 4.4.1 United-atom model for [P$_{6,6,6,14}$] cation

Compared with traditional molten salts, one fascinating feature of ILs is that they exhibit distinct heterogeneous nanostructural ordering [41]. The nanoscopic liquid organization in bulk ILs is characterized by either spongelike interpenetrating polar and apolar networks or segregated polar domains within apolar framework depending on the number of hydrophobic alkyl units in ionic species. The characteristic size of nanoscopic structural heterogeneity is found to scale linearly with aliphatic chain length in imidazolium and pyrrolidinium cations, as indicated by both SAXS experiments and atomistic simulations [36, 54, 123–127].

However, the dependence of nanoscopic liquid morphology on aliphatic chain length in tetraalkylphosphonium cation is complicated since there are four aliphatic chains in tetraalkylphosphonium cation, and each one can be tuned with varied aliphatic substituents [68, 70, 74, 75, 128] and mutated with different polar and apolar groups [34, 124]. Additionally, tetraalkylphosphonium cations can be associated with various anions and molecular liquids [33, 68, 70, 74–76, 108, 109, 129], leading to spectacular ionic structures and distinct liquid morphologies in IL matrices. The liquid organizational morphologies of [P$_{6,6,6,14}$]-based ILs, as indicated from X-ray scattering experiments and atomistic simulations conducted by Castner and Margulis groups, are dominated by three distinct landscapes at different length scales associated with short range adjacency correlations, positive–negative charge alternations at intermediate range and long-ranged polarity ordering correlations [34, 123, 124, 130].

Since aliphatic chain length is a sensitive handle to microstructural ionic environment in IL matrix, we performed atomistic simulations to elucidate the effect of linear aliphatic substituents in tetraalkylphosphonium cations on liquid landscapes, microstructures and dynamical properties of ionic species in local ionic environment [131]. Simulation results indicated that bulk tetraalkylphosphonium chloride ILs are characterized by distinct microscopic ionic structures and heterogeneous liquid morphologies depending on aliphatic chain length in tetraalkylphosphonium cations. For ILs consisting of small tetraalkylphosphonium cations, like in triethyl-butylphosphonium chloride ([P$_{2,2,2,4}$]Cl) IL, microstructural liquid morphologies are characterized by bicontinuous spongelike interpenetrating polar and apolar networks. Lengthening aliphatic chains in tetraalkylphosphonium cations leads to the polar network consisting of Cl anions and central polar groups in cations being partially broken or totally segregated within apolar framework.

The liquid morphology variations and heterogeneous microstructural changes in six tetraalkylphosphonium chloride ILs are qualitatively verified by prominent polarity alternation peaks and adjacency correlation peaks observed at low and high $q$ range in total X-ray scattering structural functions, respectively, and their peak positions gradually shift to lower $q$ values with lengthening aliphatic chains in tetraalkylphosphonium cations, as shown in Figure 4.13. The charge alternation peaks registered in intermediate $q$ range exhibit complicated dependence on aliphatic chain length in tetraalkylphosphonium cations due to the complete cancellations of positive contributions of same charge ions and negative contributions of ions of opposing charge.

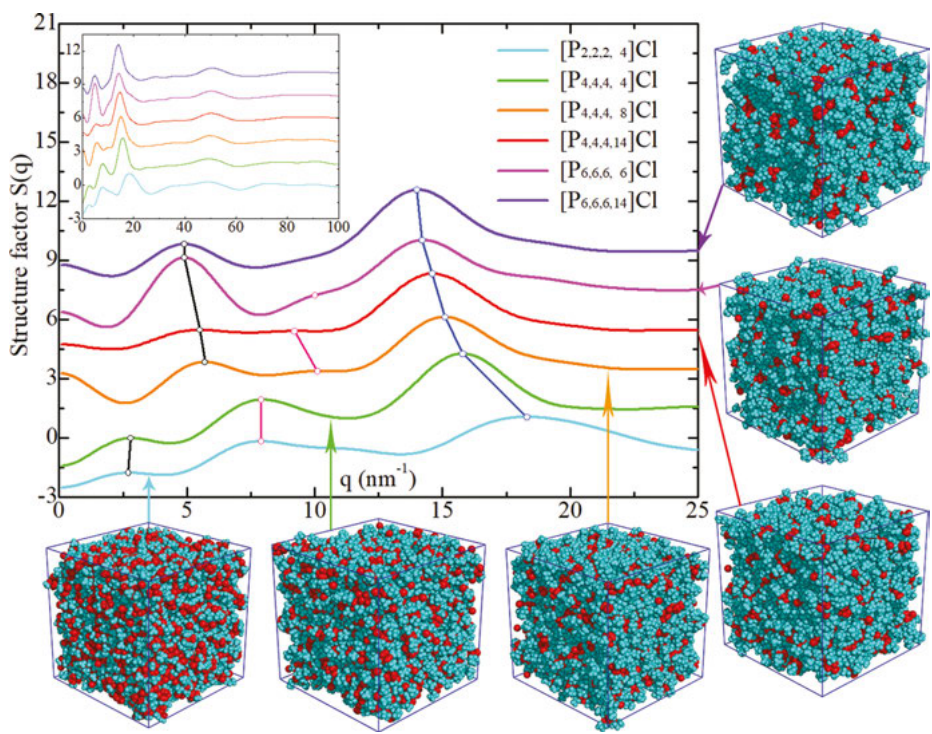

**Figure 4.13:** Structural functions $S(q)$ in the range of $q \leq 25\,\text{nm}^{-1}$ and $q \leq 100\,\text{nm}^{-1}$ (inset graph) for six tetraalkylphosphonium chloride ILs obtained from atomistic simulations at 323 K. These structural functions are vertically shifted by 2 units based on previous curves for clarity. The peak symbols in low, intermediate and high $q$ values correspond to polar–apolar alternations, positive–negative charge alternations and close contact adjacency correlations, respectively. Representative liquid morphologies of six tetraalkylphosphonium chloride ILs at 323 K are illustrated, in which polar domains (red) consist of Cl anions and central $P(CH_2)_4$ groups in tetraalkylphosphonium cations, and apolar entity (cyan) is composed of the remaining alkyl units in tetraalkylphosphonium cations, respectively.

The particular liquid morphologies and heterogeneous ionic structures in tetraalkylphosphonium chloride ILs are intrinsically manifested in dynamical properties characterized by mean square displacements, translational mobilities, van Hove correlation functions and non-Gaussian parameters of tetraalkyl-phosphonium cations and chloride anions in bulk systems [75, 100]. The terminal carbon atoms in aliphatic chains exhibit overall higher diffusivity than central phosphorus (P) atoms in tetraalkylphosphonium cations, and their cooperative effect contributes to the medium diffusion coefficients of the whole tetraalkylphosphonium cations. The P and Cl atoms exhibit comparable translational diffusivities due to their strong Coulombic coordination feature in polar domains, highlighting the existence of strongly correlated ionic structures in IL matrices. Lengthening aliphatic chains in tetraalkylphosphonium cations leads to concomitant shift of van Hove correlation functions and non-Gaussian parameters to larger radial distances and longer timescales, respectively, indicating the enhanced translational dynamical heterogeneities of tetraalkylphosphonium and chloride ions in constrained local environment.

In typical IL matrices, the liquid structural heterogeneity generally spans over an order of a few nanometers and is mainly derived from principle interactions involving different molecular moieties in ILs. The atomistic modeling of nanoscopic liquid organization of ILs over length scales beyond intermolecular distance requires that simulation system size should be several times larger than characteristic length scale of nanostructural organization of model ILs, and long time simulations should be performed to properly sample molecular conformations of ionic species in liquid organization [35, 41–44]. The latter, in particular, should be excruciatingly long due to the slow relaxation of ionic groups in bulk liquid matrix. Additionally, the lengthening aliphatic chains or increasing the number of hydrophobic alkyl units in ionic species leads to their voluminous characteristics, which further slows their reorientations in bulk region. This proposes a severe fundamental challenge for atomistic simulations despite of their initial successes in identifying the existence of nanostructural ordering phenomena in bulk ILs.

In this regard, CG models and simulations become imperative, and open a possibility to sample over large length and long time scales with a modest computational cost. The main aim of coarse-grained Molecular Dynamics (CGMD) simulations is to make this problem tractable with minimal diluting chemical rigor [35, 36, 41–43]. The CG strategy starts with a choice of specific length scale for coarsening and then subsumes all atoms presented within that length scale into one superatom or bead. These beads are then connected to one another by "bonds" to reproduce an overall architecture of CG molecule. The connected beads interact with each other via bond stretching and bending forces with similar interaction forms as that used in atomistic models but with different interaction parameters. The speedup in mimicking model IL systems using CG models is achieved, on the one hand, due to manyfold reduction in total number of degrees of freedom and the usage of softer interaction potentials, and on the other hand, due to the utilization of a larger time step than that used in atomistic simulations. These benefits make CGMD simulation a powerful way to capture slow processes that occur in complex fluids.

Due to the voluminous characteristics of tetraalkylphosphonium cations, we proposed an united-atom (UA) model for $[P_{6,6,6,14}]$ cation through a multiscale modeling protocol in which the force field parameters derived at high-resolution scale are transferred to low-resolution level in a self-consistent computational scheme using a bottom-up approach bridging different length and time scales [129]. Quantum chemical calculations were first performed to obtain the optimized molecular geometries of an isolated $[P_{6,6,6,14}]$ cation and a tightly bounded $[P_{6,6,6,14}]$ Cl ion pair structure, the latter of which is characterized by strong electrostatic interactions and moderate HB interactions between Cl anion and $[P_{6,6,6,14}]$ cation, respectively, as shown in Figure 4.14. The procedure to develop effective force field parameters and atomic partial charges for atomistic $[P_{6,6,6,14}]$ cation is the same as that used in previous works [64, 66, 68, 70]. Furthermore, an economical and nonpolarizable UA model is constructed for the $[P_{6,6,6,14}]$ cation. The hydrogen atoms in four methylene groups that are directly connected to central P atom in atomistic $[P_{6,6,6,14}]$ cation are retained in the UA model due to their preferences to form hydrogen bonds with Cl anion. Other methylene and methyl units in aliphatic chains in $[P_{6,6,6,14}]$ cation are represented as single interaction sites. The interaction parameters for UA sites are carefully tuned based on the transferable potentials for phase equilibria force field [132] to qualitatively match bond and angle distributions as that obtained from atomistic simulations, as shown in lower panels in Figure 4.14. A reduced partial charge of +0.8e is used in UA $[P_{6,6,6,14}]$ cationic model as an effective way to account for the average electrostatic polarization effect in condensed liquid state.

Atomistic and CG simulations were performed over a wide temperature range of 273–393 K to validate the proposed UA model against available experimental and computational data. The predicted volumetric quantities, including liquid density, volume expansivity and isothermal compressibility, of bulk $[P_{6,6,6,14}]$Cl IL agree well with experimental measurements. The proposed UA $[P_{6,6,6,14}]$ cationic model can essentially depict intrinsic local ionic structures and thermodynamics predicted by *ab initio* calculations and atomistic simulations, and nonlocal transport properties against corresponding experimental characterizations of $[P_{6,6,6,14}]$Cl IL over a wide temperature range. From a perspective point of view, the proposed multiscale modeling protocol to construct atomistic and UA models from *ab initio* calculations is useful and can be extended to other classes of ILs.

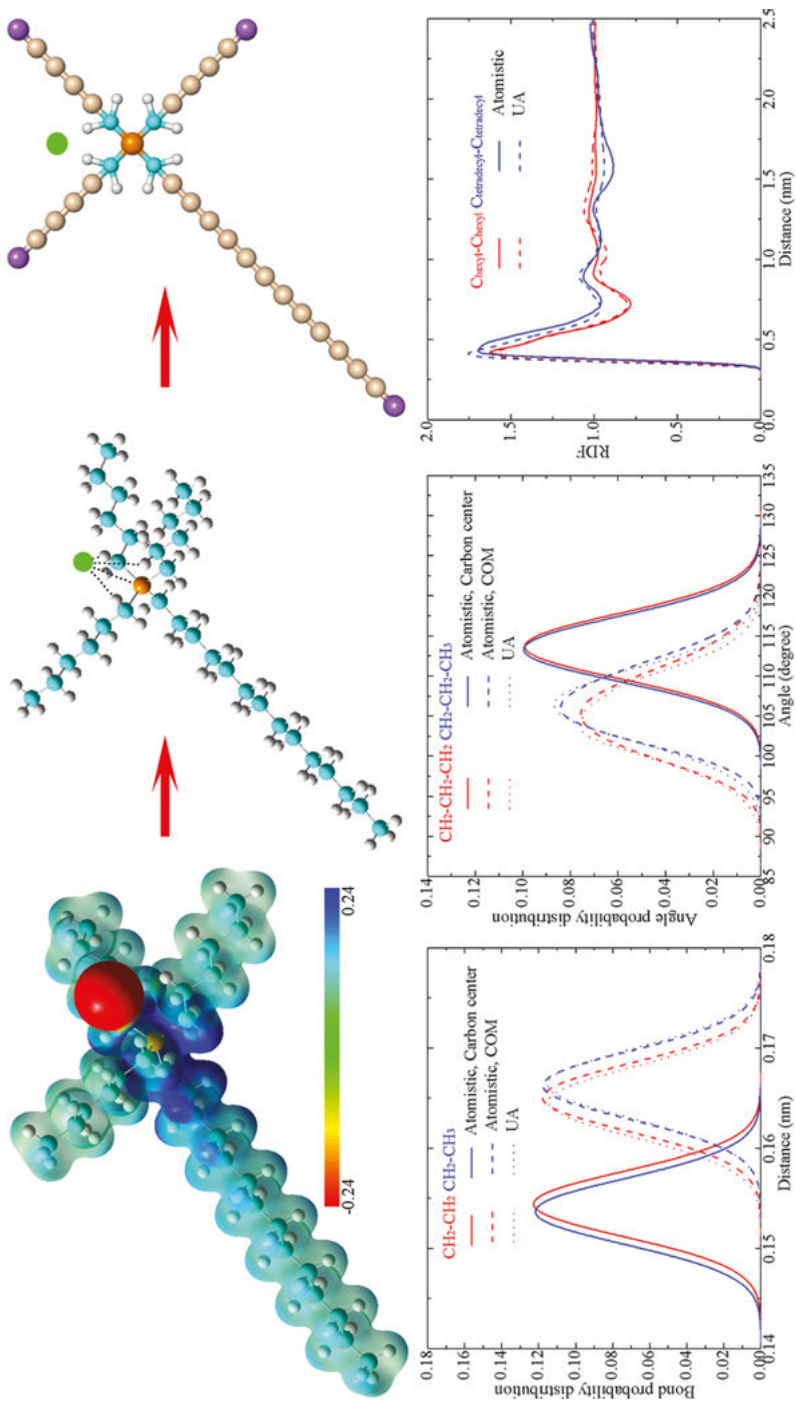

**Figure 4.14:** Molecular electrostatic potential surface and the corresponding atomistic ionic models with close contact atom pairs in a tightly bounded [$P_{6,6,6,14}$]Cl ion pair obtained from quantum chemical calculations. The bond length and angle probability distributions and radial distribution functions of terminal carbon atoms in aliphatic chains calculated from the constructed UA [$P_{6,6,6,14}$] cationic model are compared with those determined from atomistic simulations to validate proposed interaction parameters.

## 4.4.2 Coarse-grained model for [BMIM][PF$_6$] ionic liquid

By further coarse-graining of the proposed UA models to an upper level, like lumping three UA beads into a new one, one can continuously reduce computational time relative to atomistic and UA models and facilitate computational studies of solute-based dynamics on microsecond scale. Of course, this computational saving comes at an expense of some realism of real ionic models. There is an inevitable trade-off between computational cost and physical accuracy inherent to any CG model development. Additionally, CG models can reveal essential structural properties and qualitatively describe transport properties of model IL systems at mesoscopic level by integrating over less important degrees of freedom at atomic scale [31, 35, 36, 55, 56, 133, 134].

Wang and Voth first developed a generic CG model for ILs consisting of imidazolium cations coupled with [NO$_3$] anion to study their nanoscopic ionic structures with varied aliphatic chains attached to imidazolium rings [35, 36, 54, 125]. A multiscale coarse-graining approach based on the force matching method was adopted to reproduce the effective forces acting on CG beads to match with those calculated from atomistic simulations. The subsequent CGMD simulations demonstrate that charged imidazolium rings and anions organize into continuous ionic network due to strong electrostatic interactions, and the neutral aliphatic chains in cations form nonpolar domains separated from ionic framework, respectively [35, 36, 54, 125]. The geometrical constraints of head and tail groups in cations result in a novel balanced liquid crystal-like structure at low temperatures. This physical picture can qualitatively explain the experimentally observed IL crystal structure, the transition from IL to isotropic liquid crystal and changes in physicochemical and structural properties of ILs with varied aliphatic chains. In subsequent work, several coarse-graining schemes were proposed for imidazolium cations [31, 41, 55, 134]. Interaction potentials between CG beads are either obtained from specific coarse-graining approaches, like iterative Boltzmann inversion (IBI) method [134], or described by generic interaction potentials [55]. CGMD simulations show that these CG models exhibit different efficiency in describing microstructural organizations of [BMIM][PF$_6$] ILs at varied thermodynamic conditions depending on the detailed coarse-graining strategy [35–37, 41, 55, 125, 129, 134].

It should be noted that in the effective potentials derived specifically for [BMIM][PF$_6$] IL in above studies, electrostatic interactions are either explicitly calculated with Ewald summation-based methods [35, 36, 55, 125, 133], or implicitly included in interaction potentials [134]. In order to evaluate the different treatment of electrostatic interactions in CGMD simulations, we proposed a CG model for [BMIM][PF$_6$] IL based on following principles [37]: (i) Three methylene units in aliphatic chains are treated as single CG bead. This is one of the most popular coarse-graining schemes widely used in other CG simulations [55, 134]; (ii) Imidazolium ring and two remaining methyl units are symmetrically divided into two CG beads; (iii) The anionic group is described by one CG bead, which is the normal scheme in other CG models for IL system [35, 55, 133, 134]. With these guiding principles, the CG prototype for [BMIM][PF$_6$] IL was constructed and is illustrated in Figure 4.15.

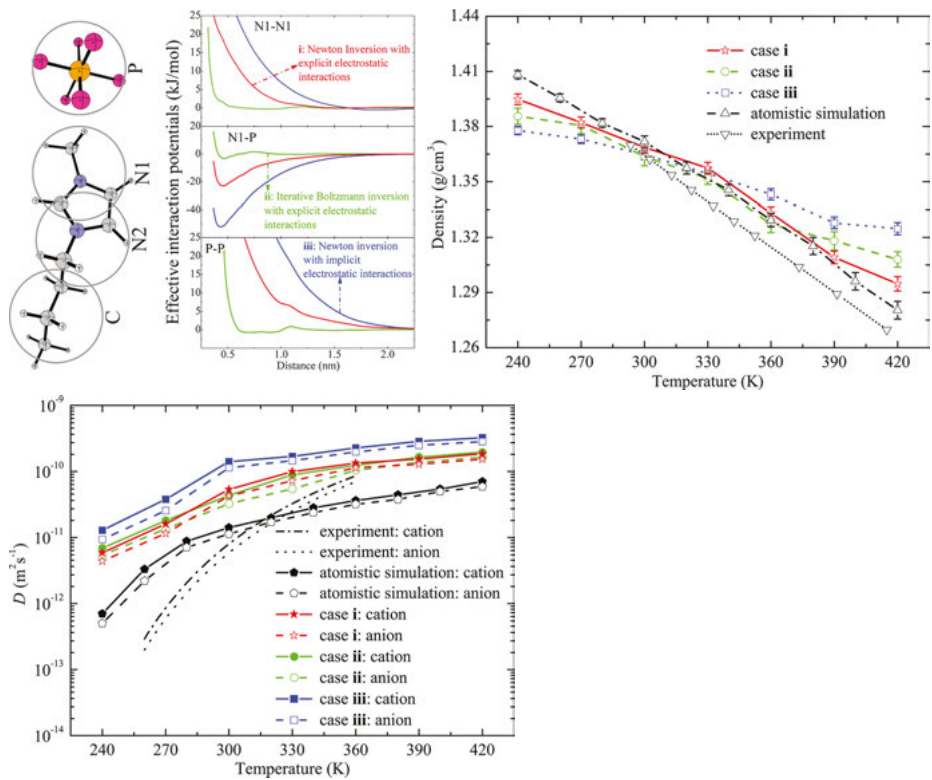

**Figure 4.15:** The CG model for [BMIM][PF$_6$] IL and typical effective interaction potentials derived from the NI and IBI methods with different treatment procedure of electrostatic interactions. Liquid densities and diffusion coefficients of [BMIM][PF$_6$] IL calculated from CG simulations using three sets of interaction potentials are compared with experimental and atomistic simulation results.

Three sets of effective interaction potentials between CG beads were derived from two iteration procedures, that is, the Newton inversion (NI) [135] and IBI methods [134], respectively, with different treatment fashion of electrostatic interactions based on radial distribution functions (RDFs) calculated from atomistic simulations. It is shown in Figure 4.15 that even the forms of effective interaction potentials derived from the NI and IBI procedures with varied treatment of electrostatic interactions are different, three sets of effective potentials can reproduce RDFs as that obtained from atomistic reference simulations with statistical uncertainty. With three sets of constructed CG potentials, we performed CG simulations on [BMIM][PF$_6$] IL over large spatial and long temporal scales and thereafter the obtained CG simulation results were comprehensively compared with experimental and atomistic simulation results on thermodynamics, microstructures, charge density distributions, scattering and dynamical properties to validate the proposed CG protocols. The interaction potentials deduced from two methods with explicit treatment of electrostatic interactions provide results that are most consistent with atomistic simulation results, whereas simulation results obtained from the effective potentials with implicit inclusion of electrostatic interactions show noticeable deviations for thermodynamic, structural and dynamical properties. In addition, the translational diffusion coefficients of ionic groups obtained from CGMD simulations are larger than that calculated from atomistic simulations, which is attributed to the simple description of [BMIM] cation and [$PF_6$] anion. This discrepancy becomes small as temperature increases, which is rationalized by the fact that force field details become unimportant and just short-ranged bead connectivity prevails at high temperatures [134]. In the development of effective potentials with the NI and IBI methods, the electrostatic interactions should be explicitly incorporated in iteration processes. In subsequent CGMD simulations, the long range electrostatic interactions should also be calculated explicitly with proper methods [136–138] to improve the reliability of dynamical properties of model IL systems at wide temperature range.

## 4.5 Conclusions and outlook

We have given several examples on how computer modeling and simulations at different length and time scales can provide valuable insight to better understand these very complex IL systems. It is clear that molecular modeling should be always closely combined with experiments as seen in the above investigations to study microstructural and dynamical properties of ILs in bulk region and in confined environment. It is fortunate that this relatively "new" class of materials has appeared at a time when simulation methodologies and computing power have converged to enable sophisticated modeling on these materials. The research field of ILs has reached an astonishing level enriched by computational knowledge gained from quantum chemical calculations, *ab initio*, atomistic and CG simulations. Had extensive interest in ILs blossomed 17 years earlier, it is unlikely that molecular simulations would be able to address all critical questions that are being tackled today. Lots of spectacular phenomena taking place at bulk region and in confined environment are still mysterious. Molecular simulations, in a long period of time in future, will be on an equal footing with experimental investigations on these striking properties of ILs in varied applications.

In future work, molecular simulations will be generally adopted in two modes within IL community. The first mode is the accurate prediction of physicochemical and microstructural quantities of model IL systems. Simulations have already been shown to make quantitative predictions of thermodynamics, microstructures, liquid morphologies and dynamics of neat ILs and IL mixtures, and this area will continue to be an important area in future investigation. The predictions of molecular properties, especially for ILs under harsh conditions where experimental characterizations are difficult to conduct, such as at high temperatures and pressures, will be of special significance. Additionally, the thermodynamic property predictions for IL mixtures, either for binary, ternary or multicomponent mixtures, should be paid more attention as these multicomponent mixtures are characterized with desirable properties that are absent in single IL sample [8, 18, 23, 34, 49]. However, care should be taken before making broad generalization. The current data set is quite restricted and some properties have been investigated for one set of ILs and other properties have been studied for others. Most IL mixtures exhibit nonideality in thermodynamics, but this does not imply that it will always be the case. Near ideal mixing is possible and has been observed in experiments [23, 49]. For IL mixtures used as absorbents for $CO_2$ capture, a critical parameter is the gas solubility of $CO_2$ in IL mixtures [19, 28]. Mixture solubility measurements are much harder to carry out experimentally than those for pure species solubility, but are in principle no more difficult to conduct in a simulation. The predicted thermodynamic properties can be further used as input parameters to test, develop and validate molecular theories, such as the statistical associating fluid theory and regular solution theory.

The second mode is providing qualitative insight into the structure–property relationship, which is even more important than calculating detailed physicochemical properties. Either IL or IL mixtures used as electrolytes in electrochemical devices [7, 17, 20, 21, 29], as absorbents for gas separation [16, 19, 28], as solvents for material synthesis or as lubricants in tribology [14, 15, 32], the overall goal is to maximize their functional performance in macroscopic applications. Clearly, such a goal is ambitious, but achievable in the long term. As discussed in previous sections, multiscale modeling simulations have already been adopted to elucidate the delicate interactions between HB and π-type interactions, to describe striking coordination pattern between residual water molecules and ionic species in IL matrices and distinct structural ordering quantities of ionic species in confinement, and to unveil nanosized polar and apolar domains within heterogeneous ionic environment. Future studies aimed to interpret the effect of molecular ionic structures on their functions in varied applications will become increasingly important. In this manner, multiscale modeling simulations may help suggest new application areas and experiments.

Despite the fact that impressive progresses have been made over the last decades, many problems still remain. Quantum chemical calculations have been invaluable in understanding the complex interplay of intermolecular forces in ILs and relationship to their thermodynamic properties. However, the proper description of transport properties is still not sufficient to generate reliable statistical data and to compute ion mobility due to limitations in computational resources. Accurate atomistic force fields must be further developed and validated for a much larger range of ionic compounds. Many force fields have been published, especially for imidazolium-based ILs [24, 46, 52, 58, 61, 62, 65, 66]. Lopes and Padua have produced a large set of force field parameters based on OPLS format [52, 67, 69], and several other groups have been active in developing force fields as well. Despite these efforts, there is a huge number of cation and anion types for which force field parameters do not yet exist. Moreover, many of the proposed force fields have not been subjected to a rigorous validation procedure. In most of these proposed force fields, liquid density is the first choice to compute and thereafter to compare with experimental measurements. This is not a good choice for force field validation since force fields having widely varying parameters can give essentially the same liquid density. This is because liquid density is essentially a mean-field property that is insensitive to specific interactions and energies [42]. It has been argued that enthalpies of vaporization, melting point and crystal structure might be better experimental quantities to compare against as these quantities are often available and extremely sensitive to the quality of force field parameters [42]. Additionally, dynamics and transport properties, which might be extremely demanding as long time simulations should be performed due to ILs' viscous feature, should be used to validate IL interaction parameters. Furthermore, as it is validated for conventional molecular liquid, one can compute liquid vapor equilibria and critical points and compare these with available experimental data. The inclusion of polarizability into constructed force fields will be an interesting topic, which can considerably increase the accuracy of the estimation of dynamics and transport properties of ILs. The computational effort for such simulations is quite large; however, with the increased computing power, the number of polarizable force fields just recently started to grow. Besides validating existing force fields, new force field parameters, or some modifications on available force field parameters, are needed for different cation and anion classes. Generating new force fields is tedious and time consuming, but must be done if molecular simulations are to be used to help guide the design of new ILs. Without new force fields that permit exploration of the diverse range of potential ILs, the relevance of molecular simulations will wane.

Another major area where opportunities for improvement exist is the development of new methods to conduct long-time simulations. CG approach is a feasible way, and it uses large time step, thereby enabling larger systems to be simulated for long times. The development of an UA model for $[P_{6,6,6,14}]$ cation and a CG model for [BMIM] cation at a higher level, as discussed in this contribution, is the typical example of this approach. It is noteworthy that the downside of CG methods is that there is an inevitable loss of some nonessential degrees of freedom at the level in which CGMD simulations are performed, and so these methods are best used when qualitative information is sought or when the desired properties are insensitive to these omitted degrees of freedom. It is shown that CG models tend to overestimate transport properties such as diffusivities. The dynamical inconsistency in CG models and its transferability are interesting topics for future research. New concepts, theories and computational tools need to be developed in the future to make truly seamless multiscale modeling a reality.

Last but not least, collaborations between simulations and experiments are crucial in order to achieve the longstanding goal of predicting particle–structure–property relationship in material design and optimization of IL systems. It is the time to carry out sophisticated computational and experimental characterizations on these materials. What could be better?

# Abbreviations

| | |
|---|---|
| ACN | acetonitrile |
| AFM | atomic force microscope |
| AIMD | *ab initio* molecular dynamics |
| B | boron |
| BF$_4$ | tetrafluoroborate |
| BMB | bis(mandelato)borate |
| BMIM | butylmethylimidazolium |
| BMLB | bis(malonato)borate |
| BOB | bis(oxalato)borate |
| BScB | bis(salicylato)borate |
| CG | coarse-grained |
| Cl | chloride |
| DFT | density functional theory |
| EAN | ethylammonium nitrate |
| EMIM | ethylmethylimidazolium |
| HB | hydrogen bonding |
| IBI | iterative Boltzmann inversion |
| IL | ionic liquid |
| LqE | low $q$ excess |
| MMIM | dimethylimidazolium |
| MS | mass spectrometry |
| NI | Newton inversion |
| NMP | N-methyl-2-pyrrolidone |
| NO$_3$ | nitrate |
| NTF$_2$ | bis(trifluoromethanesulfonyl)imide |
| O | oxygen |
| P | phosphorus |
| [P$_{2,2,2,4}$] | triethylbutylphosphonium |
| [P$_{4,4,4,8}$] | tributyloctylphosphonium |
| [P$_{6,6,6,14}$] | trihexyltetradecylphosphonium |
| PF$_6$ | hexafluorophosphate |
| PMF | potential of mean force |
| RDF | radial distribution function |
| SAXS | small-angle X-ray scattering |
| SCN | thiocyanate |
| SFA | surface force apparatus |

| SFG | sum frequency generation |
|-----|--------------------------|
| TFO | trifluoromethylsulfonate |
| TGA | thermos gravimetric analysis |
| UA | united-atom |
| vdW | van der Waals |